\documentclass[aps,pre, reprint, groupedaddress,superscriptaddress, amsmath,amssymb]{revtex4-1}

\usepackage[mathlines]{lineno}
\usepackage{color}

\usepackage{float}

\usepackage{graphicx}
\usepackage{dcolumn}
\usepackage{bm}

\begin{document}

\title{Solutal convection in porous media:\\Comparison between boundary conditions of constant concentration and constant flux}

\author{Mohammad Amin Amooie}
\email[]{amooie@mit.edu}
\affiliation{Department of Chemical Engineering, Massachusetts Institute of Technology, Cambridge, Massachusetts 02139, USA}

\author{Mohamad Reza Soltanian}
\affiliation{Department of Geology, University of Cincinnati,  Cincinnati, Ohio 45221, USA}
\affiliation{Department of Chemical and Environmental Engineering, University of Cincinnati,  Cincinnati, Ohio 45221, USA}

\author{Joachim Moortgat}
\email[]{moortgat.1@osu.edu}

\newcommand{\ext}{pdf}

\affiliation{School of Earth Sciences, The Ohio State University, Columbus, Ohio 43210, USA}

\date{\today}

\begin{abstract}
We numerically examine  solutal convection in porous media, driven by the dissolution of carbon dioxide (CO$_2$) into water---an effective  mechanism for  CO$_2$ storage in  saline aquifers. Dissolution  is associated with slow diffusion of free-phase CO$_2$ into the underlying aqueous phase followed by density-driven convective mixing of CO$_2$ throughout the water-saturated layer. We study the fluid dynamics of CO$_2$ convection in the single aqueous-phase region. A comparison is made between two different boundary conditions in the top of the formation: (\textit{i}) a constant, maximum aqueous-phase concentration of CO$_2$, and (\textit{ii}) a constant, low injection-rate of CO$_2$, such that all CO$_2$ dissolves instantly  and the system remains in single phase. {The latter model  is found to involve a nonlinear evolution  of CO$_2$ composition and associated aqueous-phase density, which depend on the formation permeability}.
We model the full nonlinear phase behavior of water-CO$_2$ mixtures in a confined domain, consider dissolution and fluid compressibility, and relax the common Boussinesq approximation. We discover new flow regimes and present quantitative scaling relations for global characteristics of spreading, mixing, and a dissolution flux in two- and three-dimensional media for both boundary conditions. We also revisit the scaling behavior of Sherwood number (Sh) with Rayleigh number (Ra), which has been under debate for porous-media convection. {Our measurements from the solutal convection in the range $1,500 \lesssim \mathrm{Ra} \lesssim 135,000$ show that the classical linear scaling Sh $\sim$ Ra is attained asymptotically for the constant-concentration case. Similarly linear scaling is recovered for the constant-flux model problem. The results provide a new perspective into how boundary conditions may affect the predictive powers of numerical models, e.g., for both the short-term and long-term dynamics of convective mixing rate and  dissolution flux  in porous media at a wide range of Rayleigh numbers.}\end{abstract}

\pacs{}

\maketitle

\section{\label{sec1}Introduction}

Convection driven by density contrast in fluids is ubiquitous in nature, and can significantly enhance the transport of  mass, heat, and energy. Examples include (thermal) convection in the Earth's mantle and atmosphere  \citep{olson1990large, cherkaoui2001laboratory}, (compositional) haline convection in sea water and groundwater aquifers  \citep{JGR:JGR11311, van2009natural}, and (thermal and compositional) double-diffusive convection in oceanic waters \citep{doubleconvection}. The latter contributes to oceanic mixing and circulation with impact on global climate. The convection process, moreover, is crucial for  successful Carbon Capture and Storage (CCS) as  one of the most promising options to stabilize atmospheric CO$_2$ concentrations and hence alleviate the global climate change  \citep{coninck2005ipcc}. Deep saline aquifers have been recognized as a primary target amongst geological formations for  CO$_2$  storage beneath the Earth's surface, where the dissolution of injected CO$_2$ into underlying water can generate convection that could help the long-term and efficient trapping of CO$_2$ \citep{szulczewski2012lifetime, keller2014potential}. How effectively convection can mix salt and thermal energy  is analogous to how effectively ``solutal convection'' in porous aquifers can mix  CO$_2$.

\begin{figure}[h]
\centerline{\includegraphics[width=.5\textwidth]{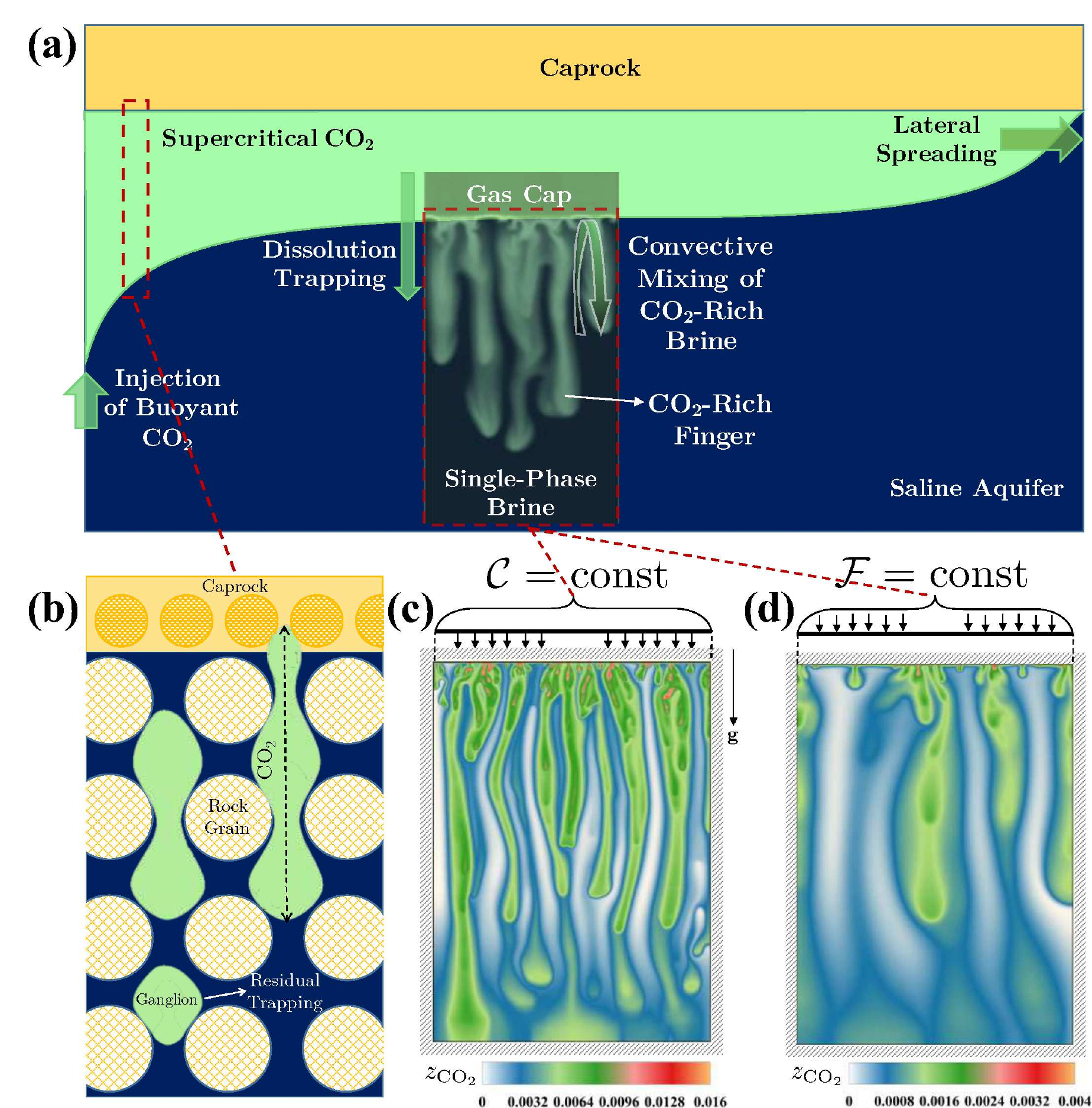}} 
\caption{Overview of {structural}, {residual}, and {dissolution trapping} mechanisms for geological storage of CO$_2$ and their relation to fluid dynamics processes such as buoyancy-driven spreading and convective mixing of CO$_2$-rich water (a). CO$_2$ rises until buoyant forces are balanced by the capillary entry pressure of the caprock (b).  The aqueous (wetting) phase displaced by CO$_2$ imbibes into pore spaces, leading to the formation of trapped CO$_2$ blobs (ganglia)---known as residual trapping \citep{gershenzon2015influence}. The single-aqueous phase in the subdomain where convection of dissolved CO$_2$ takes place is modeled under two different boundary conditions in the top: a constant-concentration (c), and a constant-flux (d). All domain boundaries are closed to flow. Snapshots in (c) and (d) are for 2D cases with $k=5,000$ mDarcy.  
\label{fig::1}}
\end{figure}

Following injection of CO$_2$ into saline formations,  buoyant (supercritical) CO$_2$ rises upward until it is confined by impermeable caprocks above the saline layer \citep{bachu2003sequestration}---known as structural trapping mechanism (Figures \ref{fig::1}a and \ref{fig::1}b). As CO$_2$ spreads laterally beneath the caprock, buoyancy poses the risk of releasing injected CO$_2$ back to the atmosphere through high-permeability pathways (e.g., faults and fractures). However, free-phase CO$_2$ gradually dissolves in the aqueous phase through diffusion, which is  referred to as \textit{dissolution trapping}  (Figures \ref{fig::1}c and \ref{fig::1}d). Over time, this mechanism can increase the storage capacity and permanence because CO$_2$ will remain in solution (even in case of caprock failure), and may eventually bind chemically to solid phases \citep{matter2009permanent, islam2016reactive, DAI2018876}. 

Dissolution of CO$_2$ into the aqueous phase creates a diffusive boundary layer that contains a fluid mixture of a higher density than the underlying fresh water. Such a density profile is gravitationally unstable, and may lead to the formation of finger-like structures (or plumes) that drive \textit{convective mixing} of CO$_2$ throughout the aquifer. Fingering is associated with the fast transport of the dissolved CO$_2$ away from the CO$_2$-water interface towards greater depths. Therefore, convection involves both diffusion of CO$_2$ from the source into the aqueous phase and the advective flow of the gravity-driven currents that carry the CO$_2$-laden water downwards. These currents simultaneously drive an upwelling flow of fresh water, thus maintaining  contact  between fresh water and source. Together, gravitational instability enhances mixing as compared to pure diffusion \citep{riaz2014carbon} and reduces the 
time-scale required for effective dissolution trapping \citep{sathaye2014constraints}. 

The convective mixing of CO$_2$ dissolved in the aqueous phase is challenging to study within the full-scale system that may consist of a two-phase (free-phase CO$_2$ and water) capillary transition zone (CTZ) between an overlying gas cap and underlying water-saturated layer \citep{emami2015two, martinez2016two}. Instead, the configuration is typically simplified to a one-phase system through one of the following assumptions:
\begin{enumerate}

\item Analogue fluid systems: in this set-up (often used in Hele-Shaw experiments), the two-phase CO$_{2}$-water system is replaced with a two-layer fluid system typically including water and a suitable fluid that is miscible with water. Fingering can be studied, but the real CO$_{2}$-water partial miscibility, density and viscosity profiles, and instability strength are only approximated \citep{backhaus2011convective, PhysRevLett.109.264503, PhysRevE.92.053023}. 
	
\item  Constant-concentration (${\cal{C}}=\mathrm{const}$) boundary condition (BC): the CO$_2$-rich layer atop the aqueous phase is replaced by a fixed impervious boundary where the solute concentration is kept at the maximum CO$_2$ solubility in water at the initial pressure ($p$)-temperature ($T$) condition \citep[e.g.,][]{pau2010high}. This model represents a canonical Rayleigh-B{\'e}nard-Darcy (RBD) problem {\citep{PhysRevLett.109.264503}}, analogous to the well-studied Rayleigh-B{\'e}nard (RB) thermal convection in free-fluid systems {\citep{pandey2018turbulent, chong2017confined}}. Multiphase processes that could affect the interface dynamics, CO$_{2}$ solubility, and associated density increases are neglected. These include the effect of interfacial tension and capillary forces within the CTZ, saturation-dependent flow constitutive relationships (e.g., relative permeability), upward penetration of water into the two-phase zone, aqueous phase volume swelling upon dissolution and the associated interface motion, pressure increases due to subsurface injection, and a drop in partial pressure of the supercritical CO$_{2}$ phase in closed systems \citep{juanes2006impact, WRCR:WRCR12159, riaz2014carbon, emami2015two, martinez2016two}. 

\item  Constant-injection (${\cal{F}}=\mathrm{const}$, or interchangeably constant-flux) BC: at a low enough injection rate (across a large interface), all CO$_{2}$ can dissolve into the aqueous phase without forming a gas cap \citep{moortgat2012three,soltanian2016critical, SoltanianAmooieEST}. The CO$_{2}$ concentration in the aqueous phase and its associated density increase  slowly in the top and then compete with the fast downward transport of CO$_{2}$ in the gravitationally unstable regime. The water density evolution is further complicated by allowing for compressibility and volume swelling of the aqueous phase (manifested by the pressure response in a confined domain) and by not adopting the Boussinesq approximation. By relaxing these limiting assumptions, interesting competitions between thermo- and hydro-dynamic processes emerge  
\citep{soltanian2016critical}. 
\end{enumerate}

The primary objective in studying dissolution trapping via natural convection is to predict the rate of CO$_2$ mixing over time. Previous experimental \citep{neufeld2010convective, backhaus2011convective, slim2013dissolution, Tsaietal} and numerical \citep{pau2010high,PhysRevLett.109.264503,  hidalgo2013dynamics, farajzadeh2013empirical, fu2013pattern, szulczewski2013carbon, slim2014solutal} studies using analogue systems and constant-concentration BC have observed a quasi-steady-state regime for both the convective flux and a mean dissipation rate. Scaling laws have been proposed for the long-term mass transport behavior in terms of Sherwood number (Sh) and Rayleigh number (Ra) (to be discussed in section \ref{sec::SH_RA}). A Sh-Ra relationship determines the ability of convection to mix the solute with ambient fluid relative to that of diffusion alone for a given buoyancy force \citep{riaz2014carbon}. Whether the dependence of Sh on Ra is linear (classical) or sublinear (anomalous) is still under debate \citep{emami2015convective}. 

In this work, we comparatively study the evolution of CO$_{2}$ mixing as well as vertical spreading for both constant-concentration and constant-injection boundary conditions, and also for both two-dimensional (2D) and three-dimensional (3D) homogeneous media. We review previous experimental and numerical studies of the long-term  behavior of natural convection, and obtain robust Sh-Ra scaling results for both model problems  through higher-order,  thermodynamically consistent  numerical simulations that account for compressibility and non-Boussinesq effects.  Our results provide new insights into the fundamental roles that phase behavior, non-Boussinesq effect, dimensionality, and boundary conditions play on solutal convection in porous media.

\section{Formulation}
\label{sec:methods}

\begin{figure*}
\centerline{\includegraphics[width=\textwidth]{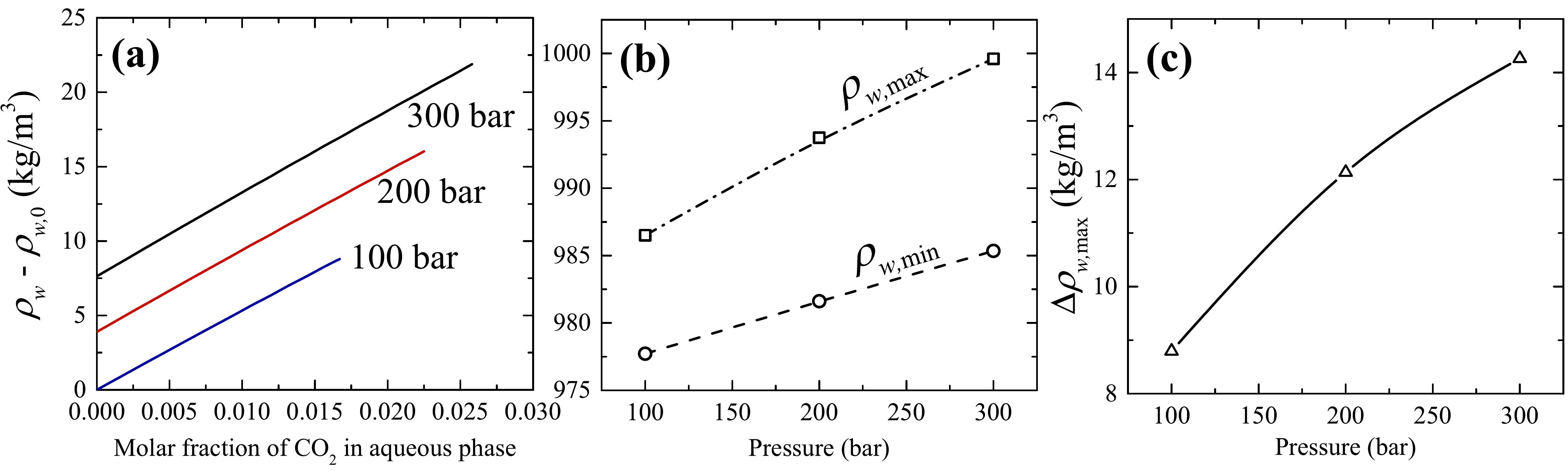}} 
\caption{{{Variation of aqueous-phase mass density as a function of pressure and molar fraction of dissolved CO$_2$. Three sample pressures (100, 200, and 300 bar) are shown. Density difference with respect to $\rho_{w,0}$, the pure water density at initial pressure (100 bar), is shown in (a). It is clear that the maximum solubility increases with pressure. The minimum ($\rho_{w,\mathrm{min}}$) and maximum density of aqueous phase ($\rho_{w,\mathrm{max}}$), corresponding respectively to zero and maximum dissolved CO$_2$ composition, are plotted in (b) as a function of pressure; the difference between the two ($\Delta\rho_{w,\mathrm{max}}$), as the main driving force to convection,  is plotted in (c) at each pressure. These results show that the density change due to dissolution is a nonlinear function of the in-situ pressure, and this should be honored.}}}
\label{fig::SI1}
 \end{figure*}

We consider inert Cartesian (vertical) 2D and 3D domains with homogeneous and isotropic permeability $k$ $\mathrm{[m^2]}$, porosity $\phi$ fields, and height $H$ $\mathrm{[m]}$. A binary mixture of CO$_2$ and H$_{2}$O is considered at isothermal conditions. {To strictly enforce mass balance at the grid cell level, we explicitly solve the molar-based conservation equations, governing transport within the aqueous phase, for both species by} 
\begin{eqnarray}
\label{eq::transfer1}\phi \frac{\partial {\cal{C}}_W}{\partial t} + \nabla\cdot \left({\cal{C}}_{W} \vec{v} +  \vec{J}_{W}\right) &=& 0, \\
\label{eq::transfer2}\phi \frac{\partial {\cal{C}}_{\mathrm{CO_2}}}{\partial t} + \nabla\cdot \left({\cal{C}}_{{\mathrm{CO_2}}} \vec{v} +  \vec{J}_{{\mathrm{CO_2}}}\right) &=& F_{\mathrm{CO_2}},
\end{eqnarray}
where ${\cal{C}}_{\mathrm{CO_2}}\equiv cz_{\mathrm{CO_2}} $ and ${\cal{C}}_W \equiv cz_W $ are each component's molar density with $c \mathrm{[mol/m^3]}={\cal{C}}_{\mathrm{CO_2}} + {{\cal{C}}}_W$ the total molar density of the mixture and $z_{\mathrm{CO_2}}$ and $z_{W}=1-z_{\mathrm{CO_2}}$  the  molar fraction of  CO$_2$ and water components, respectively. In a single phase, the phase composition of CO$_{2}$ in the aqueous phase, denoted by $x$, equals $z_{\mathrm{CO_2}}$, and short-hand notation ${\cal{C}}={\cal{C}}_{\mathrm{CO_2}}$ will be used. 
 $F_{\mathrm{CO_2}}$  $\mathrm{[mol/m^3/s]}$ is a source term for the CO$_2$ component (note that $F_W =0$ since  there is no water injection or production), $t$ is time,  $\vec{J}_{\mathrm{CO_2}}$ is the Fickian diffusive flux of CO$_2$, driven by compositional gradients \citep{moortgat2013fickian}
\begin{eqnarray}\label{eq::totmolflux}	
\vec{J}_{\mathrm{CO_2}} = -c \phi D \nabla z_{\mathrm{CO_2}}, \quad   \vec{J}_{W}=-\vec{J}_{\mathrm{CO_2}},
\end{eqnarray}
with $D=1.33\times10^{-8}\ \mathrm{m^{2}\ \mathrm{s}^{-1}}$ the constant diffusion coefficient, and $\vec{v}$ is the Darcy flux 
\begin{eqnarray}\label{Darcy}
\vec{v} = - \frac{{k}}{\mu} (\nabla p - \rho \vec{g}), 
\end{eqnarray}
with {$\vec{g} ~\mathrm{[m/s^2]}$ the gravitational acceleration}, $\mu$ $\mathrm{[kg/m/s]}$ the  phase viscosity,  and $\rho\ \mathrm{[kg/m^3] }$  the water mass density  related to the total molar density through the component molecular weights ($M$), as $\rho = {\cal{C}}_W M_W + {\cal{C}}_{\mathrm{CO_2}} M_{\mathrm{CO_2}}$. The density depends nonlinearly on not only pressure ($p$) and temperature ($T$) but also the CO$_2$  concentration, as determined by the equation of state (EOS) discussed below {(see Figure \ref{fig::SI1})}. The aqueous phase viscosity is insensitive to pressure and CO$_2$
compositions {and is assumed to only depend on temperature} $T$ (K). We use the correlation $\mu(\mathrm{cP})=0.02141 \times 10^{247.8/(T(\mathrm{K})-140)} \sim 0.3654$ \citep{moortgat2012three}. 

The Boussinesq approximation originally expresses that (\textit{i}) density fluctuations result principally from thermal effects---analogous to dissolution  here---rather than pressure effects, and (\textit{ii}) density variations are neglected except when they are coupled to gravity (i.e., in the buoyancy force, $- \rho \vec{g}$) {\citep{spiegel1960boussinesq, valori2017experimental}}. Under this approximation, density variations are small compared to velocity gradients and a divergence-free flow ($\nabla\cdot \vec{v} =0$) can be assumed. Furthermore, following an incompressible flow assumption, only a linear dependence of density on dissolved CO$_2$ concentration is typically considered (used in $- \rho \vec{g}$). In our simulations, we adopt the the full compressible and non-Boussinesq formulation by employing the cubic-plus-association (CPA) EOS---suitable for mixtures containing polar molecules---to describe the {nonlinear} dependence of density on {both} pressure and composition; density variations are also fully accounted for in both flow and transport, and the velocity field is not divergence-free ($\nabla\cdot {\cal{C}} \vec{v} \neq \vec{v} \nabla\cdot {\cal{C}}  $).  We use the same formulation as in \citet{moortgat2012three}, following \citet{li2009cubic}; for completeness the general nonlinear expressions for the EOS are provided in   Appendix \ref{appx1}. {{We also illustrate the dependence of the aqueous phase mass density on in-situ pressure and CO$_2$ composition in  Figure \ref{fig::SI1}.}} 
 
Finally, to close the system of equations, we adopt an explicit pressure equation for compressible flow based on  the \citet{acs1985general} and \citet{watts1986compositional} volume-balance approach:
\begin{align}\label{eq::acs}
\phi C_{f} \frac{\partial p}{\partial t} + \bar{\nu}_W\nabla\cdot({\cal{C}}_W\vec{v} +  \vec{J}_{W})+ \notag \\ + \bar{\nu}_{{\mathrm{CO_2}} }\left(\nabla\cdot({\cal{C}}_{\mathrm{CO_2}}\vec{v} +  \vec{J}_{{\mathrm{CO_2}}})-F_{\mathrm{CO_2}}\right) = 0,
\end{align}
where $C_f \mathrm{[Pa^{-1}]}$  is the mixture compressibility and $\bar{\nu}_i \mathrm{[m^3/mol]}$ is the  partial molar volume of each component in the mixture; both variables are computed from the CPA-EOS. 

We adopt the higher-order combination of Mixed Hybrid Finite Element and Discontinuous Galerkin methods that were presented in earlier works \citep{moortgat2011compositional, moortgat2010higher,moortgat2012three,moortgat2013CO2, moortgat2016implicit, soltanian2016simulating, GRL:GRL55656, soltanian2018impacts, soltanpfc, AMOOIE201845} for high-resolution simulations of flow and transport in porous media; more details on the numerical methods and solvers are provided in {\footnote{See Supplemental Material at [URL will be inserted by publisher] for more details on numerical methods.}}.

\section {Model Problems}

We perform 2D and 3D simulations of solutal convection in porous media. The base case 
2D domain has dimensions of 30  $\times$ 40 m$^2$, discretized by a fine 400 $\times$ 400 element mesh, and a base case 30  $\times$ 30  $\times$ 40 m$^3$   domain discretized by 90$^2$ $\times$ 100 grid is used for 3D convection. The domain size was chosen such that larger fingers are encompassed, and that the influence of boundaries on numerical solutions are minimized. {To guarantee converged results, higher grid resolutions were used for larger permeabilities (see Table I  {\footnote{See Supplemental Material at [URL will be inserted by publisher] for the grid resolutions used to achieve converged results, given in Table I.}}  for details).}  The temperature  is  77 $^{\circ}\mathrm{C}$  (170.6 $^{\circ}\mathrm{F}$). The pressure is initialized at vertical hydrostatic pressure equilibrium with 100 $\mathrm{bar}$ at the bottom.  At these conditions, the aqueous-phase density is $\rho_{w}=977.71 \ \mathrm{kg/m^{3}}$, which  increases by $\sim 0.9 \%$ (8.45 $\mathrm{kg/m^{3}}$) to $\rho = 986.16\ \mathrm{kg/m^{3}}$ when fully saturated with maximum $\sim$1.6 mol \% CO$_2$. The constant aquifer porosity is 10 \%. Homogeneous (but perturbed by a few \%) permeability fields of 250, 500, 1,000, 2,500, and 5,000 mDarcy are used in base cases. We consider bounded domains with no-flow Neumann conditions for all boundaries. The choice of no-flow, open-flow, or periodic conditions on the vertical (side) boundaries did not affect the results as long as the domains are sufficiently wide and there is no \textit{net} flux of CO$_2$ through the lateral boundaries (consistent with  \citet{juanes2006impact} and  \citet{scovazzi2013discontinuous}). 

The domain is initially saturated with fresh water (i.e., ${\cal{C}}=z_{\mathrm{CO}_2}=0$). For the constant-injection BC, CO$_2$ is introduced into the formation uniformly from top (surface in 3D) at a constant rate. This inflow is treated as source terms specified in the top-most grid cells. The injection rate is sufficiently low (0.1 \% pore volume injection, or PVI, per year), ensuring  the  CO$_2$ immediately goes into solution following the injection. That is, the CO$_2$-in-water solution thermodynamically remains under the saturation limit, maintaining a single-aqueous phase. To numerically treat the constant-concentration BC in the same framework, {we compute a source term from the diffusive flux due to the compositional gradient between the constant composition on the top edge or face and the evolving concentration at the grid center. {Therefore, both BC types are represented by  source terms that are defined in the top-most grid cells (constant for constant-injection and  variable for constant-concentration BC), as indicated in equation (\ref{eq::transfer2}).} It should be noted  that we  honor mass balance by allowing a diffusive water flux to exit the domain to satisfy the constraint $\vec{J}_{W}+\vec{J}_{\mathrm{CO_2}} = 0$ \citep{hidalgo2009effect, WRCR:WRCR12159}. We confirmed that this implementation is robust and gives similar results to another approach obtained by \citet{elenius2015interactions}, where the top-most boundary elements are initialized as the maximum molar composition and are maintained at such condition through specifying a large pore volume ($\times$ 10,000) in the top elements while reducing the  permeability by the same order (to maintain a no-flow condition across the top boundary). However, the latter approach is not as robust at high permeabilities, and the maximum concentration may still drop below the prescribed value.} 

\section{Global Characteristic Measures}
To study the general characteristics of spreading and mixing for convection, we define several quantitative global measures including (\textit{i}) dispersion-width ($\sigma_{z}$), (\textit{ii}) variance of concentration field ($\sigma^2_{{\cal{C}}}$) and individual contributions to its temporal rate, and (\textit{iii}) dissolution flux (${\cal{F}}$). Each measure is defined next.

\textit{i}) Spreading describes the average width of a spatial distribution in the mean direction of flow, and is characterized here as a longitudinal dispersion-width by the square root of the second-centered spatial variance of the CO$_2$ molar density ($\cal{C}$) in the vertical ($z$) direction \citep{aris1956dispersion,sudicky1986natural}
{\begin{eqnarray} \label{eq:dispersion}
\sigma_{z}(t) &=& 
\sqrt{\frac{\langle{\cal{C}} {z}^2\rangle}{\langle{\cal{C}}\rangle} - {\left(\frac{\langle{\cal{C}} {z}\rangle}{\langle{\cal{C}} \rangle}\right)}^2}  \!\equiv \sqrt{\left\langle\frac{{\cal{C}}}{\langle {\cal{C}} \rangle}(z-z_c)^2 \right\rangle}, 
\end{eqnarray}
where $z_c=\langle{\cal{C}} {z}\rangle/\langle{\cal{C}} \rangle$ represents  the longitudinal position of the plume center.  The notation $\langle{\cdot}\rangle$ is used for the domain averaging operator 
\begin{eqnarray} \label{eq:dispersion1}
\langle{(\cdot)}\rangle \equiv \frac{\int_\Omega (\cdot) \mathrm{d} \Omega }{\int_\Omega \mathrm{d} \Omega} =\frac{\sum_{\mathrm k\in \Omega}(\cdot)|{\mathrm k}|}{\sum_{\mathrm k\in\Omega}|{\mathrm k}|}, 
\end{eqnarray}
where $\mathrm k$ is the index of a discrete finite element (grid cell) with volume of $|{\mathrm k}|$ in medium $\Omega$. Equation (\ref{eq:dispersion}) involves  the mean square distance  from the plume centroid   in  $z$-direction {weighted} by the local probability of the CO$_2$ distribution (i.e., ${\cal{C}}/\langle {\cal{C}} \rangle$) \citep{Amooie2017}. The dispersion-width in the transverse directions is nearly constant, due to the predominantly vertical flow.	

\textit{ii}) The global variance of the CO$_2$  concentration (or molar density) field directly characterizes the mixing state of the fluid system, and is defined as
\begin{eqnarray}\label{eq::sigma2}
\sigma^2_{{\cal{C}}}(t) &=& {\langle{\cal{C}}^2\rangle - \langle{\cal{C}}\rangle^2}. 
\end{eqnarray}
The individual components that contribute to the time evolution of the domain-averaged CO$_2$ variance are linked to the fundamental character of convective \textit{mixing} and its growth rate \citep{GRL:GRL55656}. In this work, we investigate mixing for miscible, two-component, compressible transport in porous media with impermeable boundaries but subject to a CO$_2$ influx (source terms or dissolution flux) from the top boundary.  There is no mixture removal from the system, and no background flow. The goal is to derive the theoretical expressions that govern the temporal rate evolution of $\sigma^2_{{\cal{C}}}$, i.e., ${\mathrm{d}\sigma^2_{{\cal{C}} }}/{\mathrm{d} t}\equiv \dot{{\sigma^2_{{\cal{C}} }}}$. {The details of the derivations are provided in Appendix \ref{appx2} for both BCs. For the ${\cal{F}}=\mathrm{const}$ BC, we find
\begin{equation}  \label{eq::mixingRate1}
-{\phi }\frac{\mathrm{d}\sigma^2_{{\cal{C}} }}{\mathrm{d}t} =    \underbrace{-2 \bigl\langle \vec{J} \cdot \nabla{\cal{C}}  \bigr\rangle}_{2\phi\epsilon } +\underbrace{\bigl\langle {{\cal{C}} }^2\nabla\cdot \vec{v}   \bigr\rangle}_{2\phi{\cal{P}}} +\underbrace{2\bigl( \langle{\cal{C}} \rangle \langle F\rangle - \langle {\cal{C}}  F\rangle \bigr)}_{\phi \Gamma}.
\end{equation}
Equation~(\ref{eq::mixingRate1}) expresses the time evolution of the CO$_2$ global variance, and reveals the individual contributions of the mean scalar dissipation ($\epsilon$) and production (${\cal{P}}$) rates as well as the CO$_2$ source terms at the top boundary ($\Gamma$). The $\epsilon$ and ${\cal{P}}$ are analogous to those for kinetic energy dissipation and production, respectively, in turbulent flow \citep{pope2001turbulent}. 

For the ${\cal{C}}=\mathrm{const}$ BC, where CO$_{2}$ is added to the domain through a \textit{dissolution flux} along the boundary driven by \textit{diffusion},  we find
\begin{equation} \label{eq::mixingRate2}
-{\phi }\frac{\mathrm{d}\sigma^2_{{\cal{C}} }}{\mathrm{d}t} =    \underbrace{-2 \bigl\langle \vec{J} \cdot \nabla{\cal{C}}  \bigr\rangle}_{2\phi\epsilon } +\underbrace{\bigl\langle {{\cal{C}} }^2\nabla\cdot \vec{v}   \bigr\rangle}_{2\phi{\cal{P}}} +\underbrace{2{\cal{F}} \left( \langle{\cal{C}} \rangle  - {\cal{C}}_0 \right)}_{\phi \Gamma},
\end{equation}
with ${\cal{F}}$  the integrated diffusive dissolution flux across the top boundary per domain height $H$, and ${\cal{C}}_0$  the constant CO$_2$ concentration prescribed at the upper boundary.

}

\textit{iii}) The dissolution flux is a useful measure to characterize a convection process with the ${\cal{C}}=\mathrm{const}$ BC, because it defines the rate of change in the total moles of dissolved CO$_2$ within the aqueous phase per unit area. The dissolution flux is defined as
\begin{align} 
\label{eq::DissolutionFlux} 
{\cal{F}} H = \phi H \frac{\mathrm{d}\langle{\cal{C}} \rangle}{\mathrm{d}t} = \frac{H}{V}\int_{\Gamma^{\mathrm{top}}} \phi D c \nabla z_{\mathrm{CO}_2} \cdot \vec{n} \mathrm{d} \Gamma -  \notag \\ - \frac{H}{V}\int_{S} {\cal{C}}  \vec{v} \cdot \vec{n} \mathrm{d} S + H \langle F \rangle.
\end{align}
Equation (\ref{eq::DissolutionFlux}) incorporates a convective flux with respect to the vertical diffusive flux across that interface ($\sim \phi D c \nabla z_{\mathrm{CO}_2} $), the interface ($\sim {\cal{C}}\vec{v}$---applicable in two-layer or two-phase convective systems), and an injection or source term of CO$_2$ ($\langle F \rangle$). 
\begin{figure*}
\centerline{\includegraphics[width=.7\textwidth]{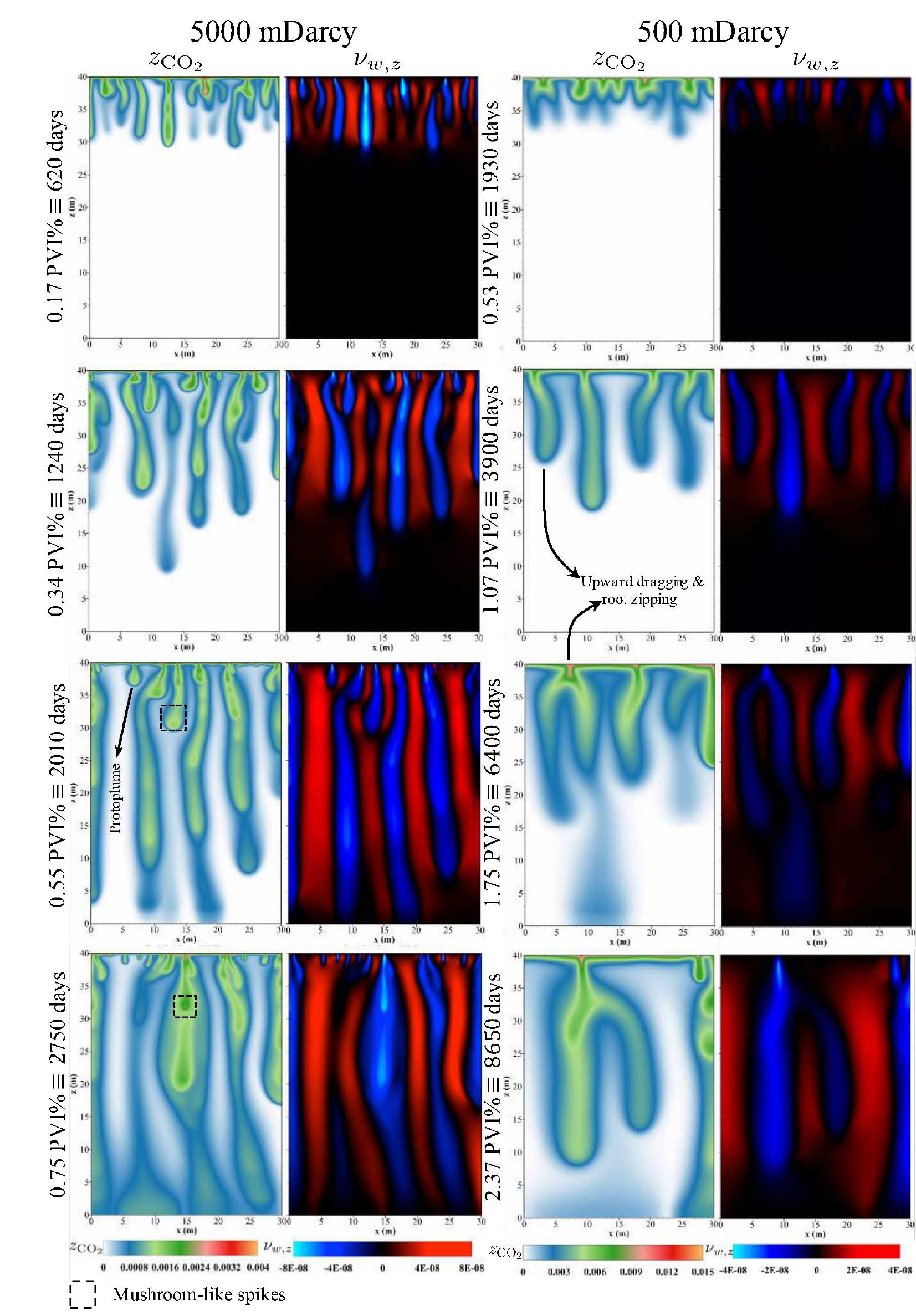}}
\caption{Constant-injection BC. Time evolution of the molar fraction of CO$_2$ ($z_{\mathrm{CO}_2}$) and the vertical Darcy velocity ($v_{w,z}$) for 5,000 (left panels) and 500 mDarcy (right panels). 
Different qualitative phenomena can be observed: downward advective flow of dense water (blue regions); reinitiation of new protoplume fingers (more pronounced in the higher permeability case) that merge with more developed megaplumes and generate mushroom-like spikes that descend; and retreating fingers that lag behind due to the upward flow generated by their faster neighbors, and subsequent root zipping. For a roughly equal front propagation in the convective regime, the lower permeability ($k_1$) case requires $\sim \sqrt{k_2/k_1}\times$ the time needed for the higher permeability ($k_2$) case. Following the advective velocity, the time for a given distance scales as $\phi \mu/kg\Delta\rho \sim k^{-0.5}$. 
}
\label{pic::1}
\end{figure*}

\begin{figure*}
\centerline{\includegraphics[width=.9\textwidth]{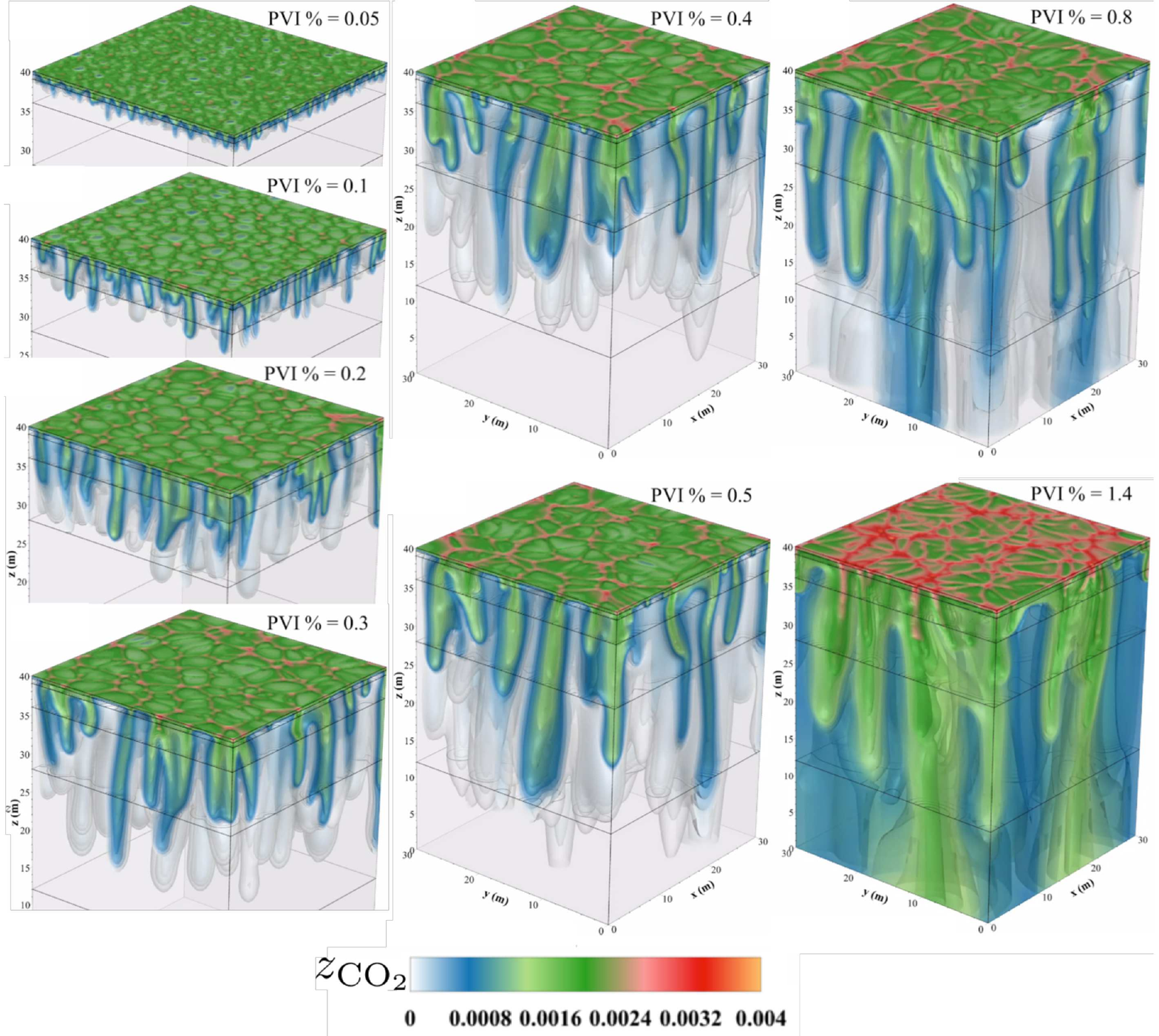}}
\caption{Snapshots of the time evolution of  CO$_2$ molar fraction in 3D convection with a constant-injection boundary condition (0.1 \% pore volume injection, or PVI, rate per year) and 5,000 mDarcy permeability.}
\label{fig::SI2}
 \end{figure*}
 
\label{sec:res}

\section{Scaling Characteristics of Spreading and Mixing Dynamics}
 
  \subsection{\boldmath{$\cal{F=\mathrm{const}}$}}

 \begin{figure*}
\centerline{\includegraphics[width=.78\textwidth]{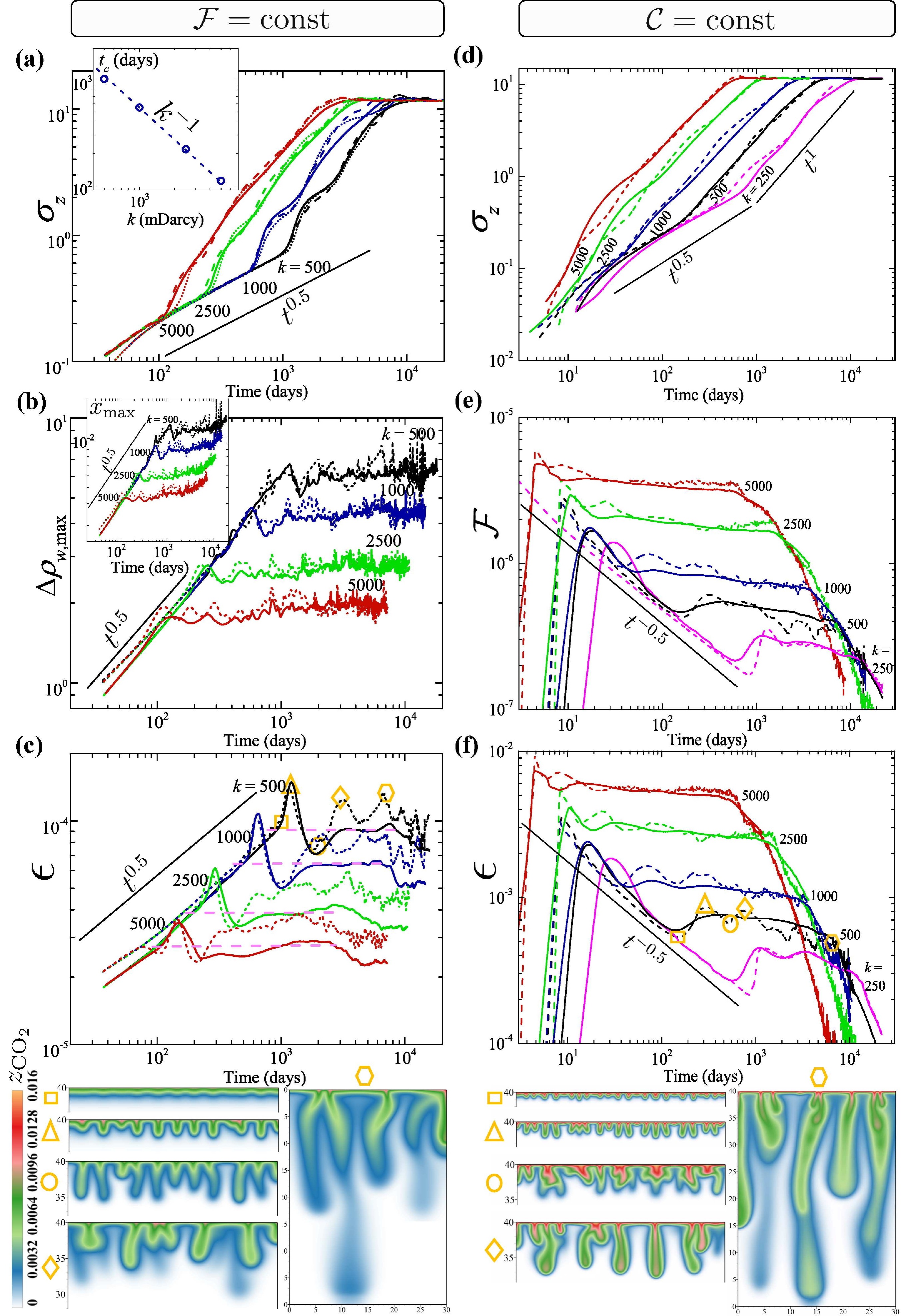}}
\caption{Quantitative characterization of CO$_2$ spreading and mixing dynamics in 2D (short-dash) and 3D (solid lines) homogeneous porous media for constant-injection or ${\cal{F}}\!=\!\mathrm{const}$ (a--c) and constant-concentration or ${\cal{C}}\!=\!\mathrm{const}$ (d--f) BC. {Dispersion-width $\sigma_z$ is shown in (a) for ${\cal{F}}\!=\!\mathrm{const}$, and  in (d) for  ${\cal{C}}\!=\!\mathrm{const}$. The time evolution of maximum density change $\Delta \rho _{w,\mathrm{max}}$ and maximum solute molar fraction $x_{\mathrm{max}}$ for the ${\cal{F}}\!=\! \mathrm{const}$ BC are shown in (b) and its inset, respectively. Mean scalar dissipation rate from global calculations, $\epsilon$, are shown in (c) and (f). The  dissolution flux per domain height for ${\cal{C}}\!=\!\mathrm{const}$ is given in (e). Results of 2D simulations with the same grid resolution as that of a vertical 2D slice through the 3D domain are plotted in dotted lines in (a), showing converged results for the 2D and 3D convection.  The key events of convective mixing from instability onset to when  fingers reach the bottom are illustrated in snapshots for $k\!=\!500$ mDarcy  in correlation with the   $\epsilon$ dynamics.}}   
\label{fig::8}
\end{figure*}

In this section we investigate the dynamical regimes of spreading and mixing of dissolved CO$_2$ in the aqueous phase for the constant-injection BC  (illustrated in Figures \ref{pic::1} as well as \ref{fig::SI2} for the 3D case with $k=5,000$ mDarcy) in terms of (\textit{i}) dispersion-width  $ \sigma_{z}$ (Figure \ref{fig::8}a), (\textit{ii}) maximum density difference between the CO$_2$-laden water and fresh water $\Delta \rho_{w,\mathrm{max}}$, and maximum molar fraction of CO$_2$ within the aqueous phase $x_{\mathrm{max}}$ (Figure \ref{fig::8}b), and  (\textit{iii}) mean scalar dissipation rate $\epsilon$ (Figure \ref{fig::8}c).

\subsubsection{Diffusive Regime}
The dispersion-width of the downward migrating plume, which is a measure of spreading, initially increases slowly at a diffusive rate as CO$_2$ is injected into the domain and thickens a diffusive boundary layer. This first period exhibits classical Fickian scaling of $ \sigma_{z} \sim t^{0.5}$, and the penetration depth scales as $\sim (Dt)^{0.5}$ \citep{einstein1905molekularkinetischen} (Figure \ref{fig::8}a). Because the concentration at the top is \textit{not} kept constant, the 
maximum density difference evolves non-trivially upon CO$_2$ dissolution (Figure \ref{fig::8}b). The temporal evolution of $\Delta \rho_{w,\mathrm{max}}$ and $x_{\mathrm{max}}$ are also Fickian, even though CO$_2$ is injected at a constant rate resulting in the linear increase of the total amount of dissolved CO$_{2}$ with time. Consistent with diffusive behavior, the time evolution of $\Delta \rho_{w,\mathrm{max}}$ and $x_{\mathrm{max}}$ in this regime are 
insensitive to permeability. 

The time evolution of the global variance rate (${\dot{\sigma^2_{{\cal{C}} }}}$), in addition to that of $\Gamma$, $\cal{P}$, and mean scalar dissipation rate from \textit{local} (grid cell) divergence values denoted now by $\epsilon^l$ (given in equation (\ref{eq::mixingRate1})) are presented for the $k=1,000$ mDarcy 2D case in Figure \ref{fig::4}a and 3D case in Figure \ref{fig::4}b. The local dissipation rate $\epsilon^l$ is more noisy in 2D than 3D, due to larger quantity of fingers overall, more surface area, and hence better numerical averaging for the integral measures in 3D, but otherwise the 2D and 3D scaling behavior is remarkably similar. 
We find that the production term is negligible, and the dynamical behavior of the variance rate
 is predominantly governed by the source of CO$_{2}$ ($\Gamma$) and its scalar dissipation rate throughout the domain. 
 
An implication of ${\cal{P}}\sim 0$ is that $2\epsilon^l \sim - {\dot{\sigma^2_{{\cal{C}} }}} - \Gamma$, {where $- {\dot{\sigma^2_{{\cal{C}} }}} - \Gamma$ is simply denoted by $2\epsilon$ for distinction in Figure \ref{fig::4}. In other words, the local dissipation rate (derived from local divergence) closely follows the indirectly computed, global one (derived by an averaging operator),  but the latter (i.e., $2\epsilon$) is obviously smoother as shown in Figure \ref{fig::4}a and in Figure \ref{fig::8}c for all the cases.} 
The absolute magnitude of these variables, given in Figure \ref{fig::4}c, demonstrate that \textit{all} the $|\Gamma|$, $2\epsilon$, and $|-{\dot{\sigma^2_{{\cal{C}} }}}|$ variables scale diffusively in this first regime but with higher absolute values for $\Gamma$ than for $2\epsilon$. This leads to a diffusive increase in the variance rate (i.e., positive ${\dot{\sigma^2_{{\cal{C}} }}}$). 

Note that $\epsilon$ (and $\Delta \rho_{w,\mathrm{max}}$ and  $x_{\mathrm{max}}$) diffusively \textit{increases} rather than \textit{decaying} as $t^{-0.5}$. The latter is the characteristic behavior for the constant-concentration BC discussed in the next section. This new behavior emerges because the diffusive decay of the concentration gradients is superimposed by a linear (in time) addition of CO$_{2}$, leading to the $\sim t^{-0.5+1=0.5}$ scaling behavior.    

 \begin{figure*}
\centerline{\includegraphics[width=1.\textwidth]{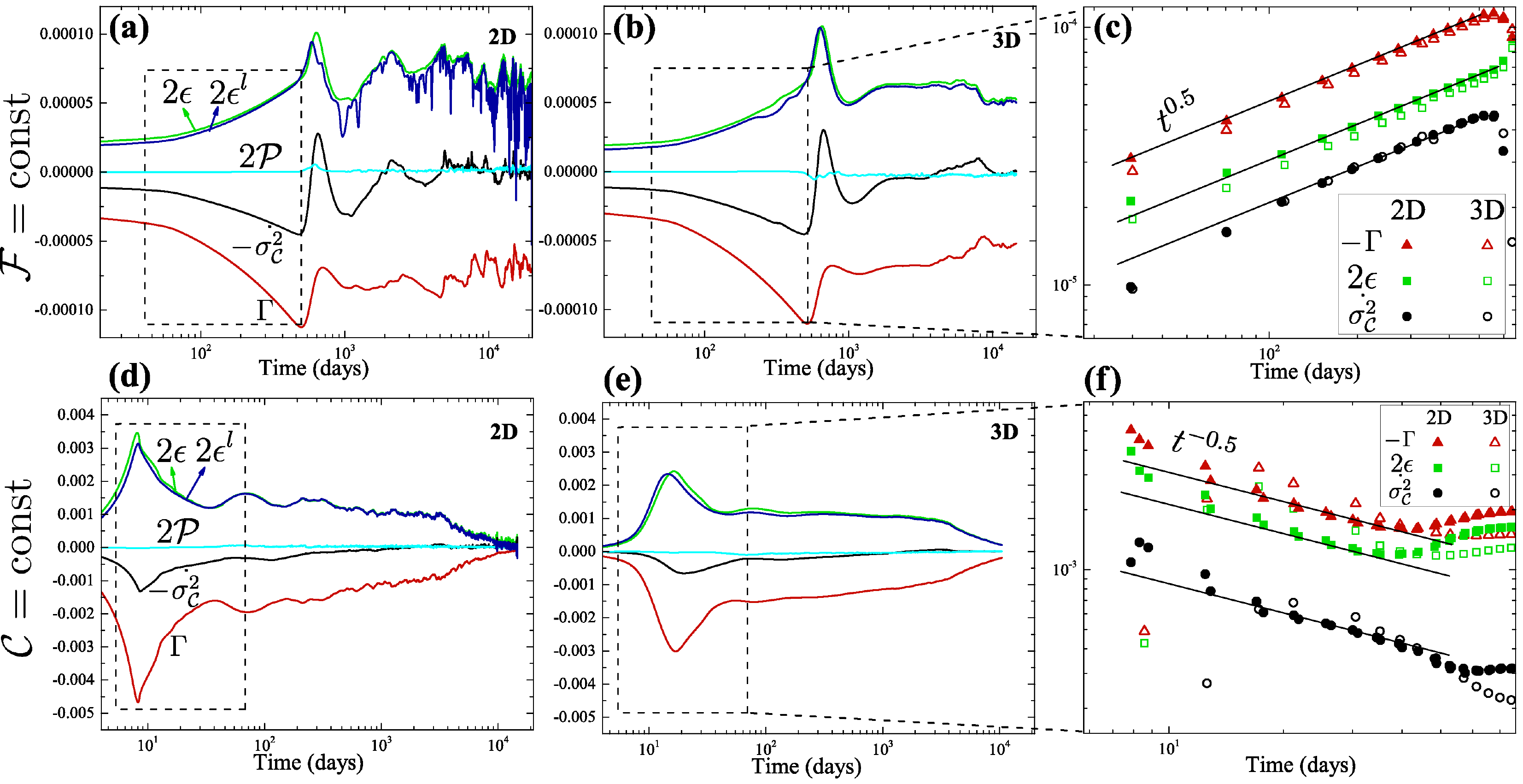}}
\caption{Evolution of temporal rate in global variance of CO$_2$ molar density, $-\dot{\sigma}^2_{{\cal{C}} }$  (with negative sign, as a proxy to global mixing rate) and the individual contributions from the mean scalar production ${\cal{P}}$ and dissipation ${{\epsilon^l}}$ rate,  obtained from local (grid cell) divergence values, as well as the contribution from the mass influx $\Gamma$. The results for 2D and 3D media with $k=1,000$ mDarcy and ${\cal{F}}\!=\!\mathrm{const}$ (respectively, ${\cal{C}}\!=\!\mathrm{const}$) are reported in (a) and (b) (respectively, (d) and (e)). ${\cal{P}}\sim 0$ implies that a less noisy dissipation rate can be estimated from global average measures (denoted here as $\epsilon$). The early evolution of absolute values for dissipation and variance rate as well as $\Gamma$ term are compared in (c) and (d) for both boundary conditions. 
}   
\label{fig::4}
\end{figure*}

\subsubsection{Early Convection} 
Density contrasts are the driving force for advective buoyant flow. For the $\cal{F=\mathrm{const}}$ BC, $\Delta \rho_{w,\mathrm{max}}$ increases slowly (diffusively) until buoyancy exceeds the diffusive restoring force and triggers a gravitational instability. This marks the onset of a flow regime where mixing eventually becomes convection dominated. All flow regimes are best captured by the evolution of $\epsilon$ (Figure \ref{fig::8}c) and the snapshots in Figure \ref{fig::8}:

\textit{i}) The departure from the diffusive scaling of $\epsilon$ occurs at the  onset of first instabilities at a critical time $t_{c}$, which exhibits a scaling relation of $k^{-1}$ for this BC (Figure \ref{fig::8}a-inset). {{This scaling of $t_{c}$ can be explained by the nonlinear evolution of densities. Linear stability analyses suggest an equation for the critical onset time as 
\begin{equation}\label{eq::tc}
t_c=c_0  \left(\frac{\phi \mu}{kg\Delta\rho} \right)^2 D = c_0\frac{1}{Ra^2}\cdot\frac{H^2}{D}, 
\end{equation}
where $c_0$ is stated to be a (numerically derived) constant and all other variables are independent from each other \citep[e.g.][]{ennis2005role, riaz2006onset, hassanzadeh2007scaling, pau2010high, slim2014solutal}. 
However, we find numerically that the maximum density increase by dissolution at the onset of instability itself is proportional to $k^{-0.5}$  (Figure \ref{fig::8}b), in line with \citep{soltanian2016critical}. More specifically, we can fit the density contrast at the critical times by  
 $\Delta \rho_{w,\mathrm{max}}(t_c) \approx \Delta \rho \left (\frac{k}{k_0} \right )^{-0.5}$ with $k_0 \approx 252~\mathrm{[mDarcy]}$  and $\Delta\rho\approx 8.45~\mathrm{[kg/m^3]}$ being the maximum density increase at the initial pressure-temperature condition. 
 Interestingly, while $t_{c} \sim k^{-1}$ and $\Delta \rho_{w,\mathrm{max}}(t_c)\sim k^{-0.5}$ follow independently from our simulations, they still satisfy equation (\ref{eq::tc}), even though the stability analyses assumed a constant density contrast.
 
 Alternatively, we can incorporate our scaling form of density difference $\Delta \rho_{w,\mathrm{max}}(t_c)\sim k^{-0.5}$  into equation (\ref{eq::tc}), and rewrite the latter in terms of independent variables but with a permeability (or Rayleigh number) dependent prefactor as:
\begin{equation}\label{eq::tc2}
t_c= \frac{k}{k_0}\cdot c_0  \left(\frac{\phi \mu}{kg\Delta \rho} \right)^2 D = \widetilde{c_0}  \left(\frac{\phi \mu}{kg\Delta \rho} \right)^2 D.
\end{equation}
This expression is interesting because it reveals consistency  with  new findings from a recent experiment \citep{rasmusson2017refractive} in which a sodium chloride (NaCl) brine solution was placed on top of and allowed to penetrate into a water-saturated silica sand box. For experimental reasons (concern of NaCl reactivity with a metal mesh at the salt-water interface), measurements were performed some distance below the actual interface, i.e., in only a subdomain inside the box unlike other studies. In this subdomain, $t_c$ was found to scale as Ra$^{-1.14}$ rather than Ra$^{-2}$ and \citet{rasmusson2017refractive} proposed a varying prefactor of $c_0 \sim Ra^{0.86}$ in relation to equation (\ref{eq::tc}) as opposed to a commonly constant prefactor. This scaling behavior is remarkably similar to our numerical findings that suggest a linear dependence. 

The reason for this different scaling in both cases is the boundary condition. In the \citet{rasmusson2017refractive} measurements,  the top of this subdomain is no longer  a no-flow boundary given the dissolved NaCl is continuously passing through it, while neither the concentration nor the concentration gradient are strictly constant across this boundary. In fact, their system of interest seems to essentially present a Robin or Dankwerts boundary condition for {transport} at the top boundary \citep{danckwerts1953continuous, brenner1962diffusion}, where the sum of advective and diffusive fluxes just below the boundary is likely constant and supplied by the stream of  solute  entering the subdomain via advection. This implies a decrease in concentration of solute from its original (saturation) limit when entering the subdomain as it undergoes the action of diffusion combined with advection. Similarly, the source term in our constant-injection BC simulations, which is simply moles per second of CO$_{2}$ entering the top grid cells, can be considered either purely advective or a sum of advective and diffusive CO$_{2}$ fluxes (although we do not consider a diffusive flux of water exiting the domain). The important implication is that CO$_{2}$ concentrations may never reach saturation levels anywhere inside the domain (e.g., when advective velocities are fast at high permeabilities). This results in the different scaling with permeabilities.

}}

Following the onset of the first instabilities, fingering generates large interfacial areas between sinking and upwelling plumes. Plume stretching simultaneously steepens the concentration gradients in the direction perpendicular to the finger \citep{GRL:GRL52302}. These mechanisms enhance mixing, and hence increase $\epsilon$ up to a global maximum. 
This `$\epsilon$-growth' regime corresponds to the first increase in dispersion-width with growing spreading rate. 

\textit{ii}) The aforementioned stretching of the CO$_2$-enriched fingers lowers the {peak} CO$_2$ composition (Figure \ref{fig::8}b-inset) at a higher rate than the replenishment of CO$_2$ from top. This causes a decrease in $\Delta \rho_{w,\mathrm{max}}$ (Figure \ref{fig::8}b) and $\epsilon$ (Figure \ref{fig::8}c) and an inflection point in $ \sigma_{z}$. A third flow regime commences in which the 
$ \sigma_{z}$ growth rate starts to decrease (Figure \ref{fig::8}a). Diffusion across the large interface between downward and upwelling plumes further decays concentration gradients. 
The \textit{negative} feedback of depleting sinking fingers of CO$_2$, and the associated $\Delta \rho_{w,\mathrm{max}}$ reduction, results in a stagnation of downward flow and stretching. 

\textit{iii}) This stagnation is the start of a fourth flow regime that is restorative. Similar to the first regime, scaling (of $\epsilon$, $\Delta \rho_{w,\mathrm{max}}$, etc.) is again approximately diffusive ($\sim t^{0.5}$) in Figures \ref{fig::8}a--\ref{fig::8}c, while the plumes become replenished by the continuous addition of CO$_2$ from the top. Coalescence and merging of slowly growing fingers lead to self-organization of fingers that cluster together to form larger-scale coherent structures.
These coarsened plumes transition into a fully developed late-convective regime once the convection driving force, $\Delta \rho_{w,\mathrm{max}}$, is \textit{restored} to exceed its value at the onset of the first instabilities. 

\subsubsection{Late Convection} 

The fifth regime is again advection (or buoyancy) dominated and displays a sharp increase in $ \sigma_{z}$ whose growth \textit{rate} is almost constant while the scaling exponents are smaller for the higher than for the lower permeability cases in this regime. The exponents are also smaller than that in the early-convection regime, consistent with findings by \citet{soltanian2016critical, SoltanianAmooieEST}. Interestingly, we discover a quasi constant-\textit{dissipation} regime for this BC, in analogy to the constant-\textit{flux} regime that is observed for the constant-concentration BC (section \ref{LateConvection}). We discuss the universality of the scaling in this regime in section \ref{SH_RA_F}.

\subsubsection{Transient Convection Shutdown}
  
Once the first fingers arrive at the bottom boundary, the dissipation rate is immediately enhanced by the mixing of laterally spreading CO$_2$-rich plume with upwelling water (Figure \ref{fig::8}c). 
As the lower boundary becomes increasingly saturated with CO$_2$, $\epsilon$ displays a late-time reduction, which characterizes a convection-shutdown regime. However, once the majority of fingers reach the bottom and undergo mixing $\epsilon$ plateaus and the shutdown regime is not persistent. This non-monotonic behavior is caused by the continuous pressure increase, and the associated increase in maximum CO$_{2}$ solubility in water (Figure \ref{fig::8}b-inset; \citep{yang2006accelerated}), as CO$_{2}$ is injected into a confined domain. Both volume swelling and fluid compressibility are taken into account in these thermodynamics effects. Following the shutdown regime, the $\sigma_{z}$ growth rate deteriorates until $ \sigma_{z}$ approaches an asymptotic value of $\sim H/\sqrt{12}$ in the limit of a spatially homogenized concentration field. 

  \subsection{\boldmath{$\cal{C=\mathrm{const}}$}}

\begin{figure*}
\centerline{\includegraphics[width=.7\textwidth]{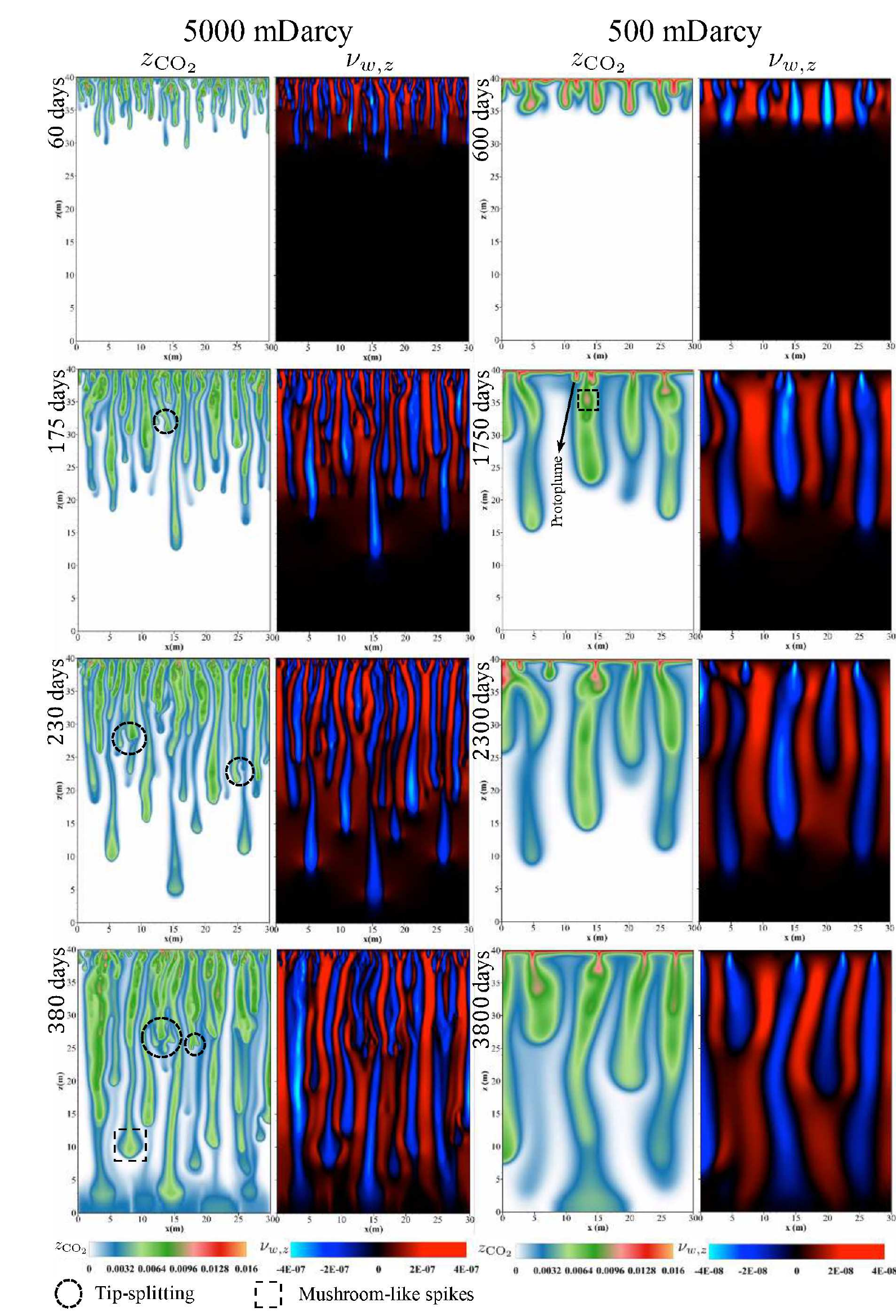}}
\caption{Constant-concentration BC. Time evolution of $z_{\mathrm{CO}_2}$ and $v_{w,z}$ for permeabilities of 5,000 (left panels) and 500 mDarcy (right panels). Different qualitative phenomena can be observed analogous to Figure~\ref{pic::1} (more pronounced here), in addition to tip-splitting in the higher permeability case.
{For a roughly equal front propagation in the convective regime, the lower permeability ($k_1$) case requires ${k_2/k_1}\times$ (5000/500 here) the time needed for the higher permeability ($k_2$) case, because the advective time scale for a given distance is proportional to $\phi\mu/kg\Delta\rho$, or $\sim k^{-1}$ for $\cal{C=\mathrm{const}}$ BC (with constant $\Delta \rho$).}
}
\label{pic::2}
\end{figure*}

  In this section, we analyze the distinct regimes in the spreading and mixing dynamics of non-Boussinesq CO$_2$ transport in the constant-concentration BC model (illustrated in Figure \ref{pic::2}), in terms of  $ \sigma_{z}$ (Figure \ref{fig::8}d), dissolution flux ${\cal{F}}$ (Figure \ref{fig::8}e), and   $\epsilon$ (Figure \ref{fig::8}f). 
   
 \subsubsection{Diffusive Regime}\label{sec::constCdiff}
   
 Similar to the constant-injection BC,  spreading and mixing are driven initially by diffusion in a growing diffusive boundary layer and $ \sigma_{z}$ again increases with classical Fickian scaling ($ \sigma_{z} \sim t^{0.5}$). However, with the diffusive transport of CO$_2$ away from the dissolution boundary, ${\cal{F}}$ and $\epsilon$ \textit{decay}  with time as $t^{-0.5}$ before the onset of instabilities, as do all the $|\Gamma|$, $2\epsilon$, and $|-{\dot{\sigma^2_{{\cal{C}} }}}|$ variables. However, we still find $|\Gamma| > 2\epsilon$ and ${\cal{P}}\sim 0$ (Figures \ref{fig::4}d--\ref{fig::4}f ). 
 {Given the step variation in the initial solute concentration (maximum at top and zero everywhere else) and assuming that the bottom boundary is sufficiently far from the top boundary during the diffusive regime, \citet{riaz2006onset} derived a 1D solution of the transport equation to describe the  evolution of concentration field within a penetrating diffusive boundary layer. The gradient of this concentration field at the top boundary, and thus the dissolution flux (see equation (\ref{eq::DisFlux}) in Appendix \ref{appx2}), follow a characteristic $t^{-0.5}$  temporal behavior \citep{riaz2006onset, slim2014solutal, martinez2016two, de2017dissolution}.}  
    
      \subsubsection{Early Convection}
   
   Once the thickness of the diffusive layer exceeds a critical value it becomes gravitationally unstable and $\sigma_{z}$, $\cal{F}$, and $\epsilon$ increase sharply in an early convection regime as compared to the diffusive regime (Figures \ref{fig::8}d--\ref{fig::8}f). {For this BC (only), $\Delta \rho_{w,\mathrm{max}}$ is constant and the onset time of the instability scales as $t_{c } \sim k^{-2} \sim Ra^{-2}$ (in dimensional form)  \citep{riaz2014carbon, emami2015convective}.}
 The early convection can be further divided into two distinct sub-regimes (also illustrated in snapshots in Figure \ref{fig::8}).
  
\textit{i}) As the dense plumes accelerate downward and fresh water is brought close to the interface, steep concentration gradients develop below the constant-concentration top boundary. In this layer, $\cal{F}$ and $\epsilon$ increase in a \textit{flux-growth} regime in analogy to the $\epsilon$-growth regime for the constant-injection BC (discussed earlier). Densely spaced fingers continue to move downward with limited lateral spreading \citep{slim2014solutal}.  

\textit{ii})  This regime of increasing $\cal{F}$-$\epsilon$ continues up to a local maximum, beyond which {merging} and shielding  between adjacent elongated fingers begin \citep{manickam1995fingering}. These interactions are promoted by diffusive spreading and the upwelling water exterior to neighboring fingers. The surviving downward `megaplumes' are more widely spaced. Concentration gradients in the boundary layer, and thus $\cal{F}$ and $\epsilon$, predominantly decrease during this \textit{merging} regime (more pronounced in 2D in Figures \ref{fig::8}e and \ref{fig::8}f).
Non-monotonic variations are caused by consecutive coalescence and growth of fingers \citep{riaz2006onset, hassanzadeh2007scaling, pau2010high, backhaus2011convective, slim2013dissolution}.

\subsubsection{Late Convection} \label{LateConvection}

While the CO$_2$ front may move faster for higher Ra cases \citep{riaz2006onset}, we find a linear growth of the global dispersion width, i.e., with ballistic  $\sim t^1$ scaling throughout the (late-time) convective regimes for all cases. 

   Finger merging continues until a quasi \textit{constant-flux} (and constant-$\epsilon$) regime develops (Figures \ref{fig::8}e and \ref{fig::8}f), analogous to the quasi constant-$\epsilon$ regime found for the constant-injection BC. While the history of events prior to this regime is different for the two different boundary conditions, the mechanisms behind the late-time behavior of convection 
 are similar and universal. In the following, we describe the long-term fate of gravitational fingers for \textit{both} BC types.
 
After the first fingers have merged and coarsened into megaplumes, and with the generation of concentration gradients below the interface due to upwelling flow of fresh water, the diffusive boundary layer thickens enough to \textit{reinitiate}
 new small-scale fingers. These features first emerge as a growing bulge on the boundary layer between the megaplumes \citep{slim2013dissolution} (Figures \ref{pic::1} and \ref{pic::2}) and are sometimes referred to as `protoplumes'.
 
The protoplumes experience three subsequent \textit{coarsening} mechanisms irrespective of BC type. 
\textit{i}) Given the impermeable top boundary, upwelling water eventually has to spread laterally and will drive nascent fingers towards the megaplumes. The protoplumes merge with the persistent megaplumes and form Rayleigh-Taylor-type mushroom spikes. These spikes can advance fast but may detach from the protoplume roots, analogous to the so-called `droplet breakup' regime in fluid mechanics \citep{Kadau08052007}. Eventually the detached CO$_{2}$ diffuses into the downwelling plumes (Figures \ref{pic::1} and \ref{pic::2}). 
\textit{ii}) Some new fingers survive and descend between the megaplumes. These features may eventually disappear either when they intersect megaplumes or through diffusive smearing. 
\textit{iii}) Some small fingers are dragged \textit{upward} by fresh water that is upwelling to accommodate the dominant megaplumes, and hence retreat as the fingers ultimately zip together from the root   (see Figure \ref{pic::1} and the animations provided in {\footnote{See Supplemental Material at [URL will be inserted by publisher] for animations.}}). 

{The consecutive events of protoplume reinitiation and coarsening establish a quasi-steady-state regime during which the boundary layer remains in a stabilizing loop: a too thin layer thickens by diffusion, while a too thick layer is stripped by the emergence and subsequent subsumption of dense protoplumes.} 

Furthermore, vigorous interactions between closely-spaced fingers, especially at high-Ra conditions and $\cal{C=\mathrm{const}}$ BC, lead to some megaplumes advancing further than others. Upwelling flow in between impacts the trailing plumes and may cause \textit{tip-splitting} in the megaplumes. When tip-splitting is followed by coarsening of those branched fingers, this can reorganize the large-scale plume structures in the interior of the domain (see Figure \ref{pic::2}). Our observations suggest that \textit{megaplumes} are not as independent from each other or persistent as previously thought \citep[e.g.,][]{slim2014solutal}. 

 The fingering interactions described above are more pronounced in higher permeability (or Ra) cases due to the denser finger population (smaller critical wavelengths). Fingering is generally more pronounced for the constant-concentration than for the constant-flux BC, because of the  smaller driving force $\sim \Delta \rho_{w,\mathrm{max}}  \sim k ^{-0.5}$ in the latter case. {As such, the difference in fingering behavior between the two BC types becomes more  pronounced as permeability increases.}

\subsubsection{Convection Shutdown}

Finally, megaplumes impact the impermeable bottom boundary, shortly after which the finite domain starts to saturate with dissolved CO$_2$---featuring again a convective shutdown regime \citep{hewitt2013convective}. $\cal{F}$ and $\epsilon$ decrease in this regime as the density (and concentration) gradients decay  in the entire domain. The shutdown regime is persistent, unlike in the constant-injection BC, because no further CO$_{2}$ will be added into the domain, but 
$ \sigma_{z}$ behaves asymptotically similar. 
     
\section{Sherwood-Rayleigh Scaling} \label{sec::SH_RA}
 Characterization of the quasi-steady-state regime is crucial to our prediction capabilities for the long-term fate and transport of CO$_2$ within saline aquifers  \citep{pruess2008numerical}.  In this section, we seek evidence of self-similar or \textit{scaling} behavior, defined  as a power-law dependence, for the evolution of the \textit{stabilized} ${\cal{F}}$ and $\epsilon$ across different media. ${\cal{F}}$ is used to obtain a Sherwood number  that characterizes the degree of convection for a given Rayleigh number. 
 
  Sh characterizes a dimensionless convective solute flux, defined as the ratio of total dissolution flux (due to advective and diffusive effects) to the purely diffusive flux:
\begin{equation} \label{eq::Sh}
\mathrm{Sh}= \frac{{\cal{F}}H}{D \phi \Delta {\cal{C}}/H} = \frac{{\cal{F}}H}{D \phi c^s_{w,\mathrm{max}} x^s_{\mathrm{max}}/H},
\end{equation}
where $\Delta {\cal{C}}= c^s_{w,\mathrm{max}} x^s_{\mathrm{max}}$ with respect to solute-free ambient fluid, with the maximum molar density $c^s_{w,\mathrm{max}}$ approximated as $ \rho^s_{w,\mathrm{max}}/(x^s_{\mathrm{max}} M_{\mathrm{CO_2}}+ (1-x^s_{\mathrm{max}}) M_W )$ and the superscript $s$ denoting the stabilized values.  Note from equation~(\ref{eq::DissolutionFlux}) that $({\cal{F}}H)$ is actually the dissolution flux across the top boundary.
Ra is a dimensionless measure that compares the time-scales of buoyancy (or natural convection) with respect to diffusive processes:
\begin{equation} \label{eq::Ra}
\mathrm{Ra}= \frac{kg\Delta\rho/\mu}{\phi D/H}.
\end{equation}
Equation~(\ref{eq::Ra}) is equivalent to the P{\'e}clet number in purely buoyancy-driven flow.  
 {$\Delta {\cal{C}}$ and $\Delta\rho$ are constant for the constant-concentration BC, with the values determined by CO$_{2}$-saturated water at the initial conditions}: $\Delta {\cal{C}}=c^s_{w,\mathrm{max}} x^s_{\mathrm{max}}\approx  855.87~\mathrm{[mol/m^3]}$, and  $\Delta\rho\approx 8.45 ~\mathrm{[kg/m^3]}$.

A classical argument requires that Sh, or the equivalent Nusselt number (Nu) for thermal convection, scale \textit{linearly} with Ra in porous-media solutal or thermal convection. The theoretical interpretation is that the flux and thus Sh in natural convection are controlled by the diffusive boundary layer, not the interior nor any external length scale. Only for an exponent of one (Sh $\sim$ Ra) does this relation become independent of $H$ \citep{howard1966convection, nield2006convection}.

We first review the recent  experimental  and numerical investigations on the Sh-Ra scaling in general convection and then discuss our own analyses.

\subsection{Experimental Studies}
\citet{Tsaietal} experimentally studied the Sh-Ra relation using water and propylene glycol (PPG) in both Hele-Shaw cells of aspect ratio one and porous media of packed glass beads in the parameter range of $10^4 \lesssim \mathrm{Ra} \lesssim 10^5$.   PPG is more dense than water, and hence represents brine while the water mimics CO$_2$ in subsurface conditions.
They obtained a scaling law of Sh $\approx0.037\mathrm{Ra}^{0.84}$. \citet{backhaus2011convective} performed experiments on the convective mass transfer with water and PPG in vertical Hele-Shaw cells of different geometric aspect ratios. A power-law relation of {{Sh $\approx(0.045\pm0.025)\mathrm{Ra}^{0.76 \pm 0.06}$}} best fitted their data for the quasi-steady regime in the parameter range of $6\times10^3 \lesssim \mathrm{Ra} \lesssim 9 \times 10^4$. 
Earlier, \citet{neufeld2010convective} developed an analogue system of methanol and ethylene-glycol (MEG)  solution and water in a porous medium (of beads). MEG is lighter than water, and hence mimics the subsurface CO$_2$. By means of a series of \textit{numerical} simulations confirming their experimental results,  \citet{neufeld2010convective} reported a power-law relationship of {{Sh $\approx(0.12\pm0.03)\mathrm{Ra}^{(0.84\pm0.02)}$}} for $2\times10^3 \lesssim \mathrm{Ra} \lesssim 6 \times 10^5 $. Based on the mixing zone model of \citet{castaing1989scaling},  \citet{neufeld2010convective} theoretically argued that the  lateral compositional diffusion from  the downward into the upwelling plumes causes the reduction of concentration as well as the driving density difference. This reduces the flux (and Sh power-law) away from the classical scaling.  While the above studies are limited to 2D convection, \citet{wang2016three} performed 3D experiments of convection in a packed bed of melamine resin particles using X-ray computed tomography. A miscible system of fluid pairs --MEG doped with sodium iodide and a sodium chloride solution-- with nonlinear  profile for mixture density was considered. A Sh $\approx0.13\mathrm{Ra}^{0.93}$ scaling was reported for a small range of  $10^3 \lesssim \mathrm{Ra} \lesssim 1.6 \times 10^4$. 

Similar non-`classical' scaling relationships have been reported in various experiments on thermal porous and free-fluid Rayleigh-B{\'e}nard convection. For instance, \citet{cherkaoui2001laboratory} performed Hele-Shaw cell heat convection experiments, and determined that  $\mathrm{Nu}\sim \mathrm{Ra}^{0.91}$ for $200 \lesssim \mathrm{Ra} \lesssim 2,000$.  High-Ra  experiments on helium gas by \citet {Heslot1987} revealed a regime of `hard turbulence' signified as $\mathrm{Nu}\approx0.23\mathrm{Ra}^{\beta=2/7}$ with $\beta$ differing from the classical 1/3 \textit{law} of natural convection in \textit{free} fluids (see discussion in \citet{otero2004high}). Sub-classical result have also been  found for different fluids  \citep{cioni1997strongly}, and phenomenologically supported by mechanistic scaling theories such as the  \citet{castaing1989scaling} mixing zone model and the \citet{ShraimanSiggia1990} nested thermal boundary layer theory.

Recently, \cite{Chingetal2017} investigated porous-media convection in Hele-Shaw cells using potassium permanganate (KMnO$_4$) powder (as CO$_2$) and water. This system of working fluids exhibits similar behavior to the CO$_2$-water system with linear increase of the mixture density due to dissolution. The experimental setup is similar to a constant-concen\-tration top BC with dissolution from the top and a linear dependence (increase) of mixture density on  dissolved KMnO$_4$, unlike the previous analogue fluid systems with nonlinear density stratification and a diffused interface between two miscible fluids shifting vertically due to volume change. They reported a \textit{linear} scaling Sh $\sim$ Ra for  $10^4 \lesssim \mathrm{Ra} \lesssim 10^6$.  

\subsection{Numerical Studies}

Several numerical studies consider convection but only a few explicitly discuss the late-time behavior. The majority of those have reported a classical linear scaling relation for the mass flux. For instance,  the 2D  simulations by \citet{pau2010high} and \citet{hesse2008mathematical}  suggest that Sh $\approx0.017 \mathrm{Ra}$ for the constant-concentration BC.  Similar results have been obtained by \citet{slim2014solutal}  for $2\times 10^3 \lesssim \mathrm{Ra} \lesssim 5 \times 10^5$, and also recovered later, in the limit of miscible convection in finite homogeneous media, using different configurations by \citet{green2014steady, de2017dissolution} (anisotropic heterogeneous media), \citet{szulczewski2013carbon}  (laterally semi-infinite domain with constant-concentration prescribed only at a finite width of the top), and \citet{elenius2014convective} and \citet{martinez2016two} (two-phase condition with CTZ). 

While all the above studies replicate the classical scaling, only two numerical studies have reported a sublinear scaling: \citet{farajzadeh2013empirical} obtained Sh $\approx0.0794\mathrm{Ra}^{0.832}$, though for a relatively {limited} range of Ra ($10^3$--$8 \times10^3$) using a constant-concentration boundary and a linear density-concentration profile; \citet{neufeld2010convective} numerically determined Sh $\approx0.12\mathrm{Ra}^{0.84}$ (also supported by experiments) for $2\times10^3 \lesssim \mathrm{Ra} \lesssim 6 \times 10^5$ but using  a mixture of two miscible fluids involving interface movement and a  non-monotonic density-concen\-tration profile. \citet{emami2015convective} concluded that the method of measuring the convective flux cannot be the source of different reported scaling behaviors. One could argue that the sublinear result of \citet{farajzadeh2013empirical} is due to the small parameter range of experiments, which includes less than  one decade of Ra. Perhaps the combination of boundary set-up and 
density-concentration profile shape determines the Sh-Ra scaling behavior, such that a constant-concentration BC with linear density-concentration profile results in linear scaling while an analogue two-layer fluid system with a non-monotonic density profile results in sublinear scaling.  
\citet{PhysRevLett.109.264503} demonstrated computationally that such an interpretation is insufficient by investigating the scaling behavior of $\epsilon$ as a proxy to the dissolution flux for the two types of models. For $5 \times10^3 \lesssim \mathrm{Ra} \lesssim 3 \times10^4$ and under the Boussinesq, incompressible fluid and miscible conditions, they showed that the stabilized $\epsilon$ exhibits no nonlinearity on Ra irrespective of the model type. 

Similar to the reviewed experiments, the nonlinear scaling behavior of heat flux (Nu) has been confirmed via numerical simulations of RB thermal porous convection. \citet{otero2004high} found a reduced exponent of $\mathrm{Nu}\approx0.0174\mathrm{Ra}^{0.9}$ for $1,300 \lesssim \mathrm{Ra} \lesssim 10^4 $. \citet{PhysRevLett.108.224503} reported a $\mathrm{Nu}\sim \mathrm{Ra}^{0.95}$ for $1,300 \lesssim \mathrm{Ra} \lesssim 4 \times 10^4 $ but suggested that the classical linear scaling is attained asymptotically (beyond Ra $\sim 10,000$).  In parallel, the 2/7 scaling for free-fluid RB convection has been also obtained via direct simulations   {\citep{kerr1996rayleigh, PhysRevLett.102.064501, van2014effect}}.

 \begin{figure}
 \centerline{\includegraphics[width=.47\textwidth]{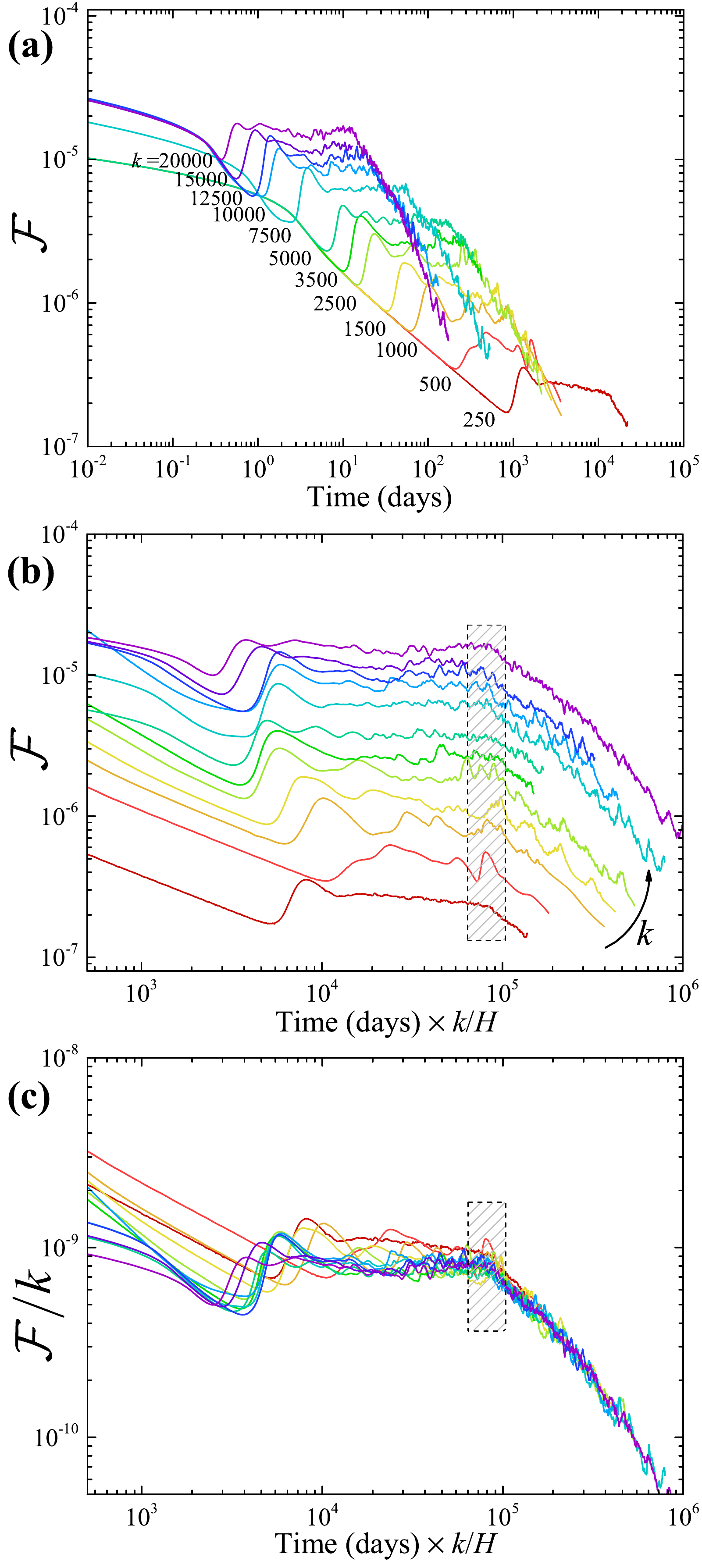}}
\caption{{{Temporal dynamics of ${\cal{F}}$, dissolution flux per domain height, in 2D convection subject to  constant-concentration  condition in the top boundary (a). Time can be rescaled by the convective time scale $\phi\mu H/kg\Delta\rho$, or simply $k/H$ provided  other parameters are constant (b). This rescaling results in approximately equal onset time of the shutdown regime following the convection regimes for different permeabilities and domains. Finally, ${\cal{F}}$ rescaled by permeability (alternatively Rayleigh number) as a function of  rescaled time shows an almost collapse of all curves in the (late) convection and shutdown periods (c). This suggests a linear Sherwood-Rayleigh scaling behavior for solutal convection is attainable.}}}
\label{fig::SI3}
 \end{figure}

 In the following, we present the Sherwood-Rayleigh scaling behavior for the problems considered in this work. 

{{
\subsection{Scaling for \boldmath{$\cal{C=\mathrm{const}}$}}
We present the results of our  high-resolution numerical simulations for 2D and 3D RBD convection in porous media.  Both dissolution flux and scalar mean dissipation rate are investigated, and  Sh-Ra scaling for a relatively wide range of Ra is reported. We extend the range of medium permeability to a maximum $k=20,000$ mDarcy (in 2D), which provides a high maximum Ra of $\sim$ 135,000 for porous media at subsurface conditions. High Rayleigh numbers
increase computational costs (higher fluxes decrease the stable time-step size) and comparison between 2D and   3D simulations was only performed up to $k=10,000$ mDarcy (i.e., Ra $\approx 67,000$). Note that the physical properties of CO$_{2}$ and water, and typical aquifer temperatures, pressures, porosity, and permeability limit the range in Ra that is meaningful in the context of CO$_{2}$ sequestration (e.g., $k=5,000$ mDarcy is already higher than typical aquifer permeabilities).

\begin{figure*}
\centerline{\includegraphics[width=.95\textwidth]{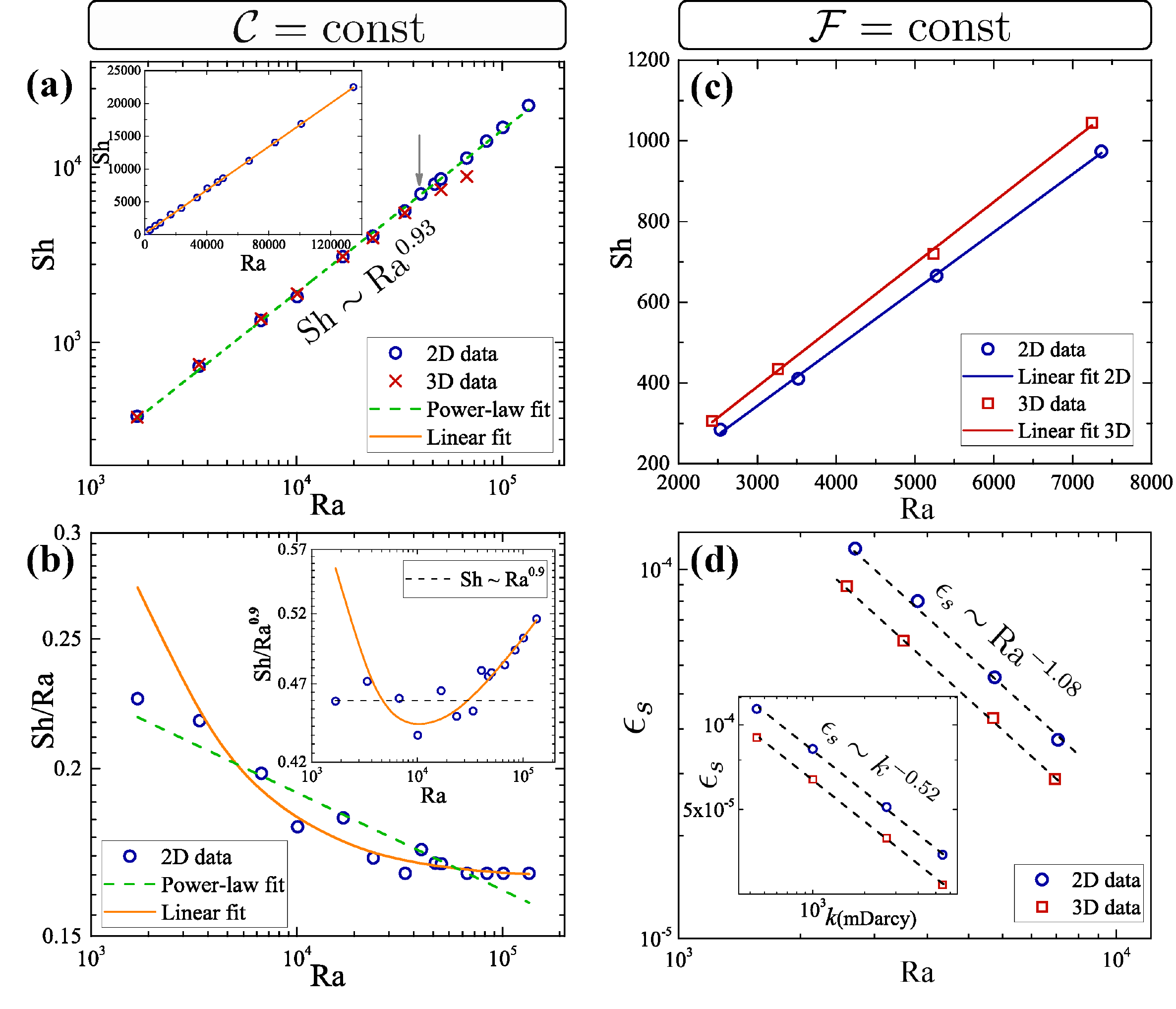}}
\caption{{{Variation of the Sherwood number Sh, a dimensionless measure of the convective flux associated with long-term convection as a function of Rayleigh number Ra. For the constant-concentration BC, Sh-Ra data for both 2D and 3D convection together with the best-fit power law scaling  are shown in (a), while the 2D data together with the best linear fit of the form  $\mathrm{Sh}\!=\!\alpha \mathrm{Ra}  + \beta$ with $\alpha\approx 0.165$ and  $\beta \approx 181.02$ are shown in the inset. Sh compensated with Ra is plotted for 2D convection in (b), together with both power-law and linear fits showing a a clear trend in Sh/Ra towards a constant as Ra increases. Sh compensated with Ra$^{0.9}$ is shown in the inset, suggesting a sublinear scaling behavior as a better fit below Ra $\approx$ 40,000 (marked by gray arrow in (a)). However, an asymptotically linear behavior  Sh $\sim$ Ra in porous-media RBD convection is concluded from (b). Linear Sh-Ra scaling recovered for 2D and 3D convection with constant-injection BC is shown in (c) and the scaling for the stabilized dissipation rate $\epsilon_s$ in (d) (inset: $\epsilon_s$-$k$ scaling).}}}  
\label{fig::6}
 \end{figure*}

 A quasi-steady regime is established in terms of both ${\cal{F}}$ and $\epsilon$ for all Ra, as shown in Figures \ref{fig::SI3}a--\ref{fig::SI3}c for ${\cal{F}}$ (as well as in  Figures \ref{fig::8}e and \ref{fig::8}f for ${\cal{F}}$ and $\epsilon$ in base cases). The 3D results exhibit less oscillations with smaller amplitude of fluctuations, which is  due to the smoother global averaging as a reflection of the additional spatial dimension over which these measures are computed.  
 Following the moving average method employed by \citet{pau2010high}, stabilized values of ${\cal{F}}$ are obtained. The latter is used to determine the strength of natural convection via Sh. We plot Sh as a function of Ra for both 2D and 3D convection in Figure \ref{fig::6}a with the least-squares power-law, and plot that for 2D convection together with  linear (i.e., first-order polynomial) fits to the measured data in Figure \ref{fig::6}a-inset. The best-fit power-law scaling for the well-validated 2D convection is Sh $\approx (0.3570\pm0.0012)Ra^{0.931\pm0.001}$ in the  range  $1,500 \lesssim \mathrm{Ra} \lesssim 135,000$ with a coefficient of determination ($R^2$) of 0.997. However, we find that the data are slightly better described by the linear fit, which takes the form
 \begin{equation} \label{eq::ShRa}
 \mathrm{Sh}=\alpha \mathrm{Ra}  + \beta; ~~\alpha\approx 0.165, ~\beta \approx 181.02,
 \end{equation} 
with a $R^2$ of greater than 0.999 over the range considered. Such scaling is suggested by Figure \ref{fig::SI3}c, which presents a better collapse of  curves in the late-convective regime for the higher permeability cases after rescaling the fluxes by $k$. Interestingly, similar scaling relations of the same form have been reported previously for 2D and 3D Rayleigh-B{\'e}nard thermal convection in a porous medium saturated with Boussinesq fluid \citep{PhysRevLett.108.224503, hewitt2014high} where there is convective transport away from {both} the upper and lower boundaries and a {statistically steady state} is attained with no shutdown period. 

We find nearly the same scaling behavior in 3D convection (for Ra $\lesssim 30,000$). Similar scaling behavior is also found for the stabilized dissipation rate in both 2D and 3D convection over the Ra range considered here (not shown), in agreement with the results of \citet{PhysRevLett.109.264503}. This is in contrast with the findings of \citet{pau2010high} (respectively, \citet{hewitt2014high}) who suggest that the 3D stabilized mass (respectively, heat) flux is typically $\sim$25\% (respectively, $\sim$40\%) {larger} than in 2D. In our simulations, the 3D scaling starts to deviate for very high Ra $> 30,000$, which could theoretically be related to increasing and more complex interactions among fingers in three dimensions, but is most likely due to numerical dispersion even when using higher-order methods on exceedingly fine grids. For all practical purposes, though, in the context of geological carbon storage, Rayleigh numbers are well below 30,000 and 2D simulations provide (surprisingly) excellent predictions for the dynamical  behavior of 3D convection.

 To shed light on the differences between the two scaling relation types (power law and linear fits) and their applicability domains, we show Sh(Ra) compensated with Ra for 2D convection in Figure \ref{fig::6}b, together with the relationship in equation (\ref{eq::ShRa}) and the power-law curve reported above. Although both scaling relations appeared to fit the data well over the full range of Ra, Figure \ref{fig::6}b reveals that a sublinear power law tends to describe the date better at lower Rayleigh numbers, while there is a clear trend in Sh/Ra towards a plateau as Ra increases beyond a transitional Rayleigh number Ra $\approx40,000$ (marked by a grey arrow in \ref{fig::6}a). This suggests that the classical linear scaling Sh $\sim$ Ra  is attained asymptotically. Next, to appreciate such distinction we show the same simulation data but rescaled by Ra$^{0.9}$ in the inset to Figure \ref{fig::6}b. A more noticeably sublinear power law Sh $\sim$ Ra$^{0.9}$ best fits the date before the aforesaid transition, while a linear fit  clearly  better represents the scaling behavior beyond that.

}}

\subsection{Scaling for \boldmath{$\cal{F=\mathrm{const}}$}}\label{SH_RA_F}

The simulations for a constant-flux BC also develop a quasi-steady-state regime and through similar governing mechanisms. 
Figures \ref{fig::8}b and \ref{fig::8}c show that   $\Delta \rho$,  $\Delta {\cal{C}}$, and the scalar dissipation rate $\epsilon$ all increase in the first convective flow regime, but then reduce and ultimately stabilize at approximately the same values as at the first onset of fingering. However, the dissolution flux ${\cal{F}}$ is now constant by definition (it is the boundary condition) and does not scale with Ra. The steady-state (stabilized) values of $\Delta \rho$ and $\Delta {\cal{C}}$ scale as $k^{-0.5}$ (Figure \ref{fig::8}b) \citep{soltanian2016critical}. Therefore, Sh $\sim (\Delta {\cal{C}})^{-1} \sim k^{0.5}$ and Ra $ \sim k \Delta \rho \sim k^{0.5}$ and thus Sh $\sim$ Ra. Specifically, Sh $\approx 0.14\mathrm{Ra}-86.9$ in 2D and Sh $\approx 0.15\mathrm{Ra}-66.5$ in 3D{, both with a  coefficient of determination of $\sim$0.999} (Figure \ref{fig::6}c).  Similar to $\Delta \rho$,  $\Delta {\cal{C}}$, the stabilized dissipation rate ($\epsilon_s$) approximately scales as $\epsilon_s \sim k^{-0.52}$, as shown in the inset to Figure \ref{fig::6}d.
 This is consistent with the observations that $\epsilon \sim t^{0.5}$ in the first diffusive regime, and $t_c\sim k^{-1}$ (Figure~\ref{fig::8}a-inset), and thus $\epsilon_s \sim k^{-0.5} \sim Ra^{-1}$ (as observed in Figure \ref{fig::6}d).

The physical reason that the Sh-Ra scaling for the constant-composition BC shows more complex behavior could be a feedback loop between the {supply} of new CO$_{2}$ (${\cal{F}}$) and the flow dynamics inside the domain. Conversely, for a constant-flux BC, convection is fully determined by the properties inside the domain (e.g., permeability). We also point out that the driving force for convection ($\Delta \rho_{w,\mathrm{max}}$) is stronger in the constant-composition BC, which shows more pronounced fingering. This may explain why the constant-flux BC simulations, where the maximum driving force is inversely proportional to permeability,  do not  show an increase in tip-splitting and transverse finger interactions at high Ra.
 
\section{Discussion and Concluding Remarks}

We analyze detailed simulations in 2D and 3D of gravity-driven natural convection of a solute, specifically CO$_{2}$ dissolved in water, in deep subsurface porous aquifers. Our results are an improvement over earlier studies both in terms of numerical methods and physical assumptions. Higher-order finite element methods and fine grids are used to fully resolve the small-scale fingering and tip-splitting. The commonly used Boussinesq approximation is relaxed, and we allow for (molar) density gradients in flow and transport equations, in addition to fluid compressibility, volume swelling, and other thermodynamic phase behavior effects through an accurate equation of state (CPA-EOS). 
Other novel findings follow from a detailed comparison between different boundary conditions in the top of the domain: the common constant-composition BC and a constant-flux  BC in which CO$_{2}$ is injected at a low rate such that the water remains under-saturated.

For both BC, we study the {global} evolution of spreading (dispersion-width) and mixing (mean scalar dissipation rate) of CO$_{2}$. We also compare this to the evolution of the \textit{locally} derived individual contributions to the mixing rate. The latter analysis suggests that compressibility and non-Boussinesq effects do not significantly impact spreading and mixing. 

Both BC models develop a quasi-steady-state following the early-time convection and before the shutdown regime in response to new plume nucleation balancing the merger between earlier plumes. For the constant-concentration BC, the quasi-steady-state is usually expressed as a plateau in the dissolution flux, but this definition is not applicable in the constant (dissolution) flux BC. Instead, one can use the plateau in mean scalar dissipation rate to define the quasi-steady-state regime, as it can be applied to both BC for characterizing the dynamical behavior of convective mixing.

{{

 Particular attention is paid to how the Sherwood number in the quasi-steady-state regime scales with the Rayleigh number. For the constant-concentration BC model, the nature of such relationship  has been the subject of recent debate.  Our scaling analyses reveal that the measurements of the convective flux over the range $1,500 \lesssim \mathrm{Ra} \lesssim 135,000$ are best fitted by an expression of the form Sh $=\alpha \mathrm{Ra} + \beta$ with $\alpha\approx 0.165$ and $\beta\approx 181.02$. Particularly, such linear fit performs better than  the best-fitted power law Sh $\approx (0.3570\pm0.0012)\mathrm{Ra}^{0.931\pm0.001}$  beyond  Ra $\approx40,000$. This suggests that the classical linear scaling is attained asymptotically, even in non-Boussinesq, compressible model of convective mixing, and that the previously reported sublinear relations could be in part a result of relatively limited parameter range of experiments below an asymptotic limit.

}}

For the case of a constant-injection BC, the dissolution flux is constant by definition. However, we show that the maximum density and concentration change evolve dynamically in time, rather than being imposed as constants, against the rate at which the dissolved CO$_2$ migrates downwards. Furthermore, they become stabilized in correlation with the dynamics of mixing rate, while all scaling as $\sim k^{-0.5}$. These relations recover the classical linear Sh-Ra scaling for this boundary condition. 

The scaling relations and analyses of convection dynamics developed in this work have a broad applicability to other density-driven problems such as mantle convection \citep{olson1990large}, oceanic circulations, atmospheric convection \citep{cherkaoui2001laboratory}, and haline convection in sea water \citep{JGR:JGR11311} and groundwater aquifers \citep{van2009natural}. Convection dynamics for the constant-injection BC can be applied to examples of constant-flux water infiltration into a porous medium resulting in gravity-driven fingering \citep{PhysRevLett.101.244504}, thermal convection with a constant heat flux at top and bottom boundaries \citep{PhysRevLett.102.064501}, the saltwater bucket problem \citep{WRCR:WRCR12159}, and the proposed injection of CO$_2$-saturated water into saline aquifers \citep{lindeberg2003long}.

 \appendix
\section{Cubic-plus-association equation of state}\label{appx1}

Phase behavior is obtained from the CPA-EOS, which honors the thermophysical aspects of CO$_2$-water mixtures and is able to accurately reproduce measured densities as well as partial molar volumes (for the swelling effect). This is unlike most previous studies that relied on simplified linear or empirical correlations for  mixture density and Henry's law for $\mathrm{CO}_2$ solubilities \citep[e.g.,][]{farajzadeh2013empirical}. CPA-EOS is an improvement over cubic EOS for fluid mixtures that contain polar molecules such as water. Through thermodynamic perturbation theory, it takes into account all the polar-polar interactions including the self-association of water molecules and (polarity-induced)  cross-association between water and CO$_2$ molecules \citep{li2009cubic, moortgat2012three, firoozabadi2015thermodynamics}. We use the same CPA formulation as in \citet{moortgat2012three}, following \citet{li2009cubic}.

Similar to the ideal gas law, molar density is related to pressure as $c=p/ZRT$ with $R$ the universal gas constant.  $Z$ is the compressibility factor, that accounts for the nonideal behavior of fluid, i.e., all the polar-polar interactions. $Z$  primarily depends  on $T$, $p$, and $z_{\mathrm{CO_2}}$ as well as the critical properties and binary interaction coefficients (BICs) of water and CO$_2$, expressed as follows:
\begin{widetext}
\begin{eqnarray}\nonumber
\label{eq::CPA}
&Z = \underbrace{\frac{Z}{Z-B}-\frac{AZ}{Z^2 + 2BZ -B^2}}_{\mathrm{physical}} +\underbrace{\frac{4+4\eta-2\eta^2}{2-3\eta+\eta^2}  \left[z_W(y_W-1)+ z_{\mathrm{CO_2}}(y_{\mathrm{CO_2}}-1) \right]}_{\mathrm{association}}, & \\\nonumber\\\nonumber
&\mathrm{with} \quad \eta=\frac{B}{4Z}, \quad y_W=\frac{Z}{Z+2z_W y_W  \delta + 2 z_{\mathrm{CO_2}} y_{\mathrm{CO_2}} s \delta}, \quad  y_{\mathrm{CO_2}}=\frac{Z}{Z+2z_W y_W s \delta},& \\\nonumber\\
&\mathrm{where} \quad \delta=\frac{1-0.5\eta}{(1-\eta)^3}\frac{\xi p}{RT}\left[\mathrm{exp}\left(\frac{\epsilon}{k_B T} \right) -1\right], \quad s=0.0529T^2_r+0.0404T_r-0.0693.&
\end{eqnarray}
\end{widetext}

 $A$ and $B$ (respectively, $\epsilon$ and $\xi$) are respectively bonding energy and volume parameters of physical  interactions (respectively, association).   The $A$ and $B$ can be estimated by applying the van der Waals quadratic mixing rules and proper BICs. $k_B$ is the Boltzmann constant. $y_W$ and $y_{\mathrm{CO_2}}$ denote respectively the mole fractions of water and CO$_2$ molecules that are not bonded at one of the association sites. $\delta$  represents the association strength between water molecules while $s \delta$ is the association between water and CO$_2$ molecules with $s$ the cross association factor. $T_r=T/T_c$ is the reduced  temperature of CO$_2$ with $T_c$ the critical temperature of CO$_2$. 
 
 {
 \section{Detailed derivation of equations for global variance evolution}\label{appx2}

We derive the theoretical expressions that govern the temporal rate evolution of $\sigma^2_{{\cal{C}}}(t) = {\langle{\cal{C}}^2\rangle - \langle{\cal{C}}\rangle^2}$, i.e., ${\mathrm{d}\sigma^2_{{\cal{C}} }}/{\mathrm{d} t}\equiv \dot{{\sigma^2_{{\cal{C}} }}}${, following previous analyses of mixing in viscously unstable flows \citep{jha2011quantifying, jha2011fluid, nicolaides2015impact, GRL:GRL55656}}. Multiplying equation (\ref{eq::transfer2}) by ${{\cal{C}}}$, we obtain
\begin{eqnarray}
\label{eq::transfer_times}\phi {\cal{C}} \frac{\partial {\cal{C}}}{\partial t} + {\cal{C}}\nabla\cdot \left({\cal{C}}\vec{v} +  \vec{J}\right) = {\cal{C}}F,
\end{eqnarray}
where ${\cal{C}}\nabla\cdot \left({\cal{C}}\vec{v}\right)$ and ${\cal{C}}\nabla\cdot  \vec{J}$ can be respectively expanded as $\frac{1}{2}{\cal{C}}^2\nabla \cdot \vec{v} +\frac{1}{2}\nabla \cdot\left({\cal{C}}^2\vec{v} \right)$ and $\nabla\cdot({\cal{C}}\vec{J})-\vec{J}\cdot\nabla{\cal{C}}$. Depending on the top BC, ${\cal{F}}=\mathrm{const}$ or ${\cal{C}}=\mathrm{const}$, the derivation of ${\mathrm{d}\sigma^2_{{\cal{C}} }}/{\mathrm{d} t}$ is different. 

For the ${\cal{F}}=\mathrm{const}$ BC: Applying the Gauss divergence theorem to the bounded domain, one obtains $\langle \nabla \cdot \left({\cal{C}}^2\vec{v} \right) \rangle=\langle \nabla \cdot ({\cal{C}}\vec{J} ) \rangle=0$ (injection term appears as source term $F$). Therefore, volume averaging equation (\ref{eq::transfer_times}) yields
\begin{eqnarray}\label{eq::transfshort}
\phi      \frac{\mathrm{d} \langle{{\cal{C}}}^2 \rangle}{\mathrm{d} t}  =   2 \left\langle \vec{J} \cdot \nabla{\cal{C}} \right\rangle -\left\langle {{\cal{C}}}^2\nabla\cdot \vec{v}   \right\rangle +2\langle {\cal{C}} F\rangle.
\end{eqnarray}

Similarly, by integrating equation~(\ref{eq::transfer2}) over the domain and then applying the divergence theorem, we find $\mathrm{d}\langle{\cal{C}} \rangle/\mathrm{d}t=\langle F\rangle/{\phi}$.
Writing the rate of change in equation~(\ref{eq::sigma2}) as
\begin{eqnarray}
\frac{\mathrm{d}\sigma^2_{{\cal{C}} }}{\mathrm{d} t}= \frac{\mathrm{d}\langle{\cal{C}} ^2\rangle}{\mathrm{d}t}-2\langle{\cal{C}} \rangle \frac{\mathrm{d}\langle{\cal{C}} \rangle}{\mathrm{d}t},
\end{eqnarray}
and combining all the above terms, we finally find
\begin{equation}  \label{eq::mixingRate1_apx}
  \boxed{
-{\phi }\frac{\mathrm{d}\sigma^2_{{\cal{C}} }}{\mathrm{d}t} =    \underbrace{-2 \bigl\langle \vec{J} \cdot \nabla{\cal{C}}  \bigr\rangle}_{2\phi\epsilon } +\underbrace{\bigl\langle {{\cal{C}} }^2\nabla\cdot \vec{v}   \bigr\rangle}_{2\phi{\cal{P}}} +\underbrace{2\bigl( \langle{\cal{C}} \rangle \langle F\rangle - \langle {\cal{C}}  F\rangle \bigr)}_{\phi \Gamma}.}
\end{equation}

For the ${\cal{C}}=\mathrm{const}$ BC: CO$_{2}$ is added to the domain through a \textit{dissolution flux} along the boundary driven by \textit{diffusion}. Therefore, $\langle \nabla \cdot ({\cal{C}}\vec{J} ) \rangle \neq 0$ while $\langle \nabla \cdot \left({\cal{C}}^2\vec{v} \right) \rangle=F=0$. The equation for the mean concentration is obtained by integrating equation~(\ref{eq::transfer2}), which yields $\mathrm{d}\langle{\cal{C}} \rangle/\mathrm{d}t=-\langle \nabla\cdot  \vec{J}\rangle/\phi$.  Using the Gauss divergence theorem gives
\begin{align} \label{eq::DisFlux}
\langle \nabla\cdot  \vec{J}\rangle= \frac{1}{V}\int_{S} \vec{J} \cdot \vec{n} \mathrm{d} S= \frac{1}{V}\int_{\Gamma^{\mathrm{top}}} \vec{J} \cdot \vec{n} \mathrm{d} \Gamma =  \notag \\ = -\underbrace{\frac{1}{V}\int_{\Gamma^{\mathrm{top}}} \phi D c \nabla z_{\mathrm{CO}_2} \cdot \vec{n} \mathrm{d} \Gamma}_{{\cal{F}}}\quad \Rightarrow
\frac{\mathrm{d}\langle{\cal{C}} \rangle}{\mathrm{d}t}=\frac{{\cal{F}}}{\phi}
\end{align}
with $S$ denoting the full surface (and $\mathrm{d} S$ its increment) of the domain with volume $V$, and $\Gamma^{\mathrm{top}}$ (with increment $\mathrm{d} \Gamma$) is the surface of the top boundary, with $\vec{n}$ the corresponding outward-pointing normal ($z$ increases downward from $z=0$ in the top).
 ${\cal{F}}$ is the integrated diffusive dissolution flux across the top boundary (i.e., $  -\frac{1}{A}\int_{\Gamma^{\mathrm{top}}} \phi D c \frac{\partial z_{\mathrm{CO}_2}}{\partial z}   \mathrm{d} \Gamma$) per domain height $H$. 
 We also have $\langle \nabla \cdot ({\cal{C}}\vec{J} ) \rangle   = -{\cal{C}}_0 {\cal{F}}$, because the CO$_2$ concentration is a {constant} ${\cal{C}}_0$ at the upper boundary. Finally, we obtain an expression analogous to equation (\ref{eq::mixingRate1_apx}) but now for the constant-concentration BC
\begin{equation} \label{eq::mixingRate2_apx}
  \boxed{
-{\phi }\frac{\mathrm{d}\sigma^2_{{\cal{C}} }}{\mathrm{d}t} =    \underbrace{-2 \bigl\langle \vec{J} \cdot \nabla{\cal{C}}  \bigr\rangle}_{2\phi\epsilon } +\underbrace{\bigl\langle {{\cal{C}} }^2\nabla\cdot \vec{v}   \bigr\rangle}_{2\phi{\cal{P}}} +\underbrace{2{\cal{F}} \left( \langle{\cal{C}} \rangle  - {\cal{C}}_0 \right)}_{\phi \Gamma}.}
\end{equation}
}
 
 \begin{acknowledgments}
The authors greatly thank Juan J. Hidalgo, Joaqu{\'\i}n Jim{\'e}nez-Mart{\'\i}nez, and anonymous reviewers for useful  discussions and comments. Acknowledgment is made to the Donors of the American Chemical Society Petroleum Research Fund for partial support of this research.
 The second author was supported by the U. S. Department of Energy's (DOE) Office of Fossil Energy funding to Oak Ridge
National Laboratory (ORNL) under project FEAA-045. ORNL is managed by UT-Battelle for the U.S. DOE under Contract
DE-AC05-00OR22725.
All data and methodology are provided in the main text, Supplemental Material, and references.

\end{acknowledgments}

 \end{document}